\documentclass[aps,prd,onecolumn,showpacs,superscriptaddress,nofootinbib,preprintnumbers]{revtex4-2}

\usepackage{style}

\newcommand{\bea}{\begin{equation}\begin{aligned}}
\newcommand{\eea}{\end{aligned}\end{equation}}

\begin{document}


\title{\textbf{The Fate of Large Scale White Noise in\\ Second-Order Cosmological Perturbation Theory}}

\author{Aurora Ireland\,\orcidlink{0000-0001-5393-0971}} 
\email{anireland@stanford.edu}
\affiliation{Leinweber Institute for Theoretical Physics, Stanford University, Stanford, CA 94305}

\date{\today}

\begin{abstract}
Working to second order in cosmological  perturbation theory, we reconsider the Large Scale White Noise (LSWN) effect proposed in Refs.~\cite{LSWN,Noisy}. We demonstrate that the kurvature density variable $\Delta \rho$ does indeed develop LSWN at second order. However, contrary to the claims of~\cite{Noisy}, we show that this is not inherited as an infrared (IR) pole in the second-order comoving curvature perturbation $\mathcal{R}_2$. This apparent enhancement was an artifact of using a linear Poisson equation to relate two genuinely second order quantities. Further, we show that $\mathcal{R}_2$ is protected from developing any such IR enhancement: $\lim_{k \rightarrow 0} \left( k^2 \mathcal{R}_2 \right) = 0$, which follows from a conservation law in the soft limit. The constraint proposed in~\cite{Noisy} and its implications for the small-scale primordial power spectrum are therefore invalid.
\end{abstract}

\maketitle
\section{Introduction}

The primordial curvature perturbation $\mathcal{R}$ is a central object in early-universe cosmology. Generated during inflation, it provides the initial conditions for the subsequent evolution of cosmological perturbations. At leading order, its statistics are characterized by the power spectrum $\mathcal{P}_\mathcal{R}(k)$. On large scales, this is constrained primarily through the cosmic microwave background (CMB) temperature and polarization anisotropies. On smaller scales, weaker constraints from large-scale structure and various other late-time observables exist. On the smallest scales, however, $\mathcal{P}_\mathcal{R}(k)$ remains essentially unconstrained. Any observable sensitive to this regime would therefore provide a new window onto the early universe. 

A different measure of the curvature inhomogeneity was recently introduced in Refs.~\cite{LSWN,Noisy}\footnote{Of which the present author is a co-author.} in connection with this otherwise unconstrained regime. For a fluid with local energy density $\rho$ and 4-velocity $u^\mu$, the so-called ``kurvature density'' is defined as \bea\label{eq:kurvature:def}
    \Delta \rho = \rho - \frac{\theta^2}{24 \pi G} \,,
\eea
where $\theta = \nabla_\mu u^\mu$ is the local expansion rate of the fluid. This combination measures the mismatch between the amount of matter present in a region and how rapidly that region is expanding. In a spatially flat homogeneous FLRW spacetime it is vanishing, and so $\Delta \rho$ measures the local deviation from flatness. Unlike the usual variables of cosmological perturbation theory, $\Delta \rho$ is defined covariantly, without reference to a choice of coordinates or perturbative gauge, and is built from quantities measurable in the fluid rest frame. At linear order, $\Delta \rho$ is directly proportional to the intrinsic spatial curvature of comoving hypersurfaces---and hence, through a Poisson equation, to $\nabla^2 \mathcal{R}$ (see Eq.~(\ref{eq:kurvature:curvature:1st:order:relation})). This last relation is what motivates its name, and it is what Ref.~\cite{Noisy} used to convert statements about $\Delta \rho$ into statements about $\mathcal{R}$.

It was observed in Refs.~\cite{LSWN,Noisy} that beyond linear order, this quantity is sourced by the quadratic combination $\sigma^2 - \omega^2$ built from the shear $\sigma_{\mu\nu}$ and vorticity $\omega_{\mu \nu}$ of the fluid. Since these terms carry no spatial gradients, it was anticipated that the spectrum of $\Delta \rho$ should be white noise on large scales, i.e.
\bea\label{eq:white:noise:def}
    \lim_{k \rightarrow 0} P_{\Delta \rho}(k) = \text{constant} \,.
\eea
To understand the origin of this expectation, note that a quantity sourced quadratically acquires at second order a convolution over primordial perturbations of the form $\int d^3 q \, f(\vec{q}\,) f(\vec{k} - \vec{q}\,)$. When $f$ carries no positive powers of external momentum $k$, this integral tends towards a constant in the soft limit $k \rightarrow 0$,
\bea\label{eq:convolution:integral}
    \lim_{k \rightarrow 0} \int d^3 q \, f(\vec{q}\,) f(\vec{k} - \vec{q}\,) \sim \int dq \, q^2 f(q)^2 = \text{constant} \,,
\eea
presuming the kernel is such that the integral is bounded. By contrast, quantities whose sources involve spatial gradients acquire compensating powers of $k$ and so are suppressed in this limit. 

That the quantity $\Delta \rho$ should develop this large scale white noise (LSWN) is worth emphasizing because it is not what one typically expects. Most of the usual quantities of interest in cosmology obey some sort of local conservation law which protects them from developing LSWN. The canonical example is the density contrast $\delta$, for which mass and momentum conservation enforce $P_\delta(k) \sim k^4$ on large scales~\cite{Peebles:1974,Peebles:1980}. The kurvature density evades these sorts of constraints since it has no corresponding conservation laws. 

If the LSWN in $\Delta \rho$ were somehow propagated to large-scale observables, it would give a rare window onto small-scale primordial power, since the convolution integral in Eq.~(\ref{eq:convolution:integral}) receives contributions from arbitrarily high internal loop momenta $q$. This would be remarkable. The CMB and large-scale structure together tightly constrain $\mathcal{P}_\mathcal{R}(k)$ only for $k \lesssim 1 \, \text{Mpc}^{-1}$. Beyond this, one is limited to much weaker and more indirect probes, such as CMB spectral distortions, scalar-induced gravitational waves, and primordial black hole abundances. A mechanism capable of transferring power from arbitrarily small scales to the largest observable scales would sidestep these entirely, turning the best measured observable in cosmology into a probe of its least accessible regime. 

This was exactly what was claimed in Ref.~\cite{Noisy}, wherein the LSWN in $\Delta \rho$ was converted into a statement about the comoving curvature perturbation $\mathcal{R}$. Using the linear Poisson relation $\nabla^2 \mathcal{R}_1 = 4 \pi G a^2 \Delta \rho_1$ extrapolated to second order, \cite{Noisy} argued that $\mathcal{R}$ should inherit a $1/k^2$ infrared (IR) enhancement in the soft limit---leading to a contribution to the (dimensionless) power spectrum of the form $\mathcal{P}_\mathcal{R} \supset k_{\rm BH}/k$. By demanding that this contribution not exceed the observed power in CMB anisotropies at small angular multipoles, \cite{Noisy} was able to bound the wavenumber $k_{\rm BH}$. Because $k_{\rm BH}$ was computed as an integral over the primordial power spectrum of the form $k_{\rm BH} \sim \int d\ln q \mathcal{W}(q) \mathcal{P}_\mathcal{R}^2(q)$, the bound on $k_{\rm BH}$ in turn was inferred to constrain the allowed size of the small-scale power spectrum. 

Crucially, however, this procedure is not self-consistent. In particular, while this Poisson-type relation is exact at linear order, at second order $\Delta \rho$ acquires extrinsic curvature contributions encoded in the quadratic form $\mathcal{Q}$, while $\mathcal{R}$ requires a quadratic completion $\mathcal{C}$, such that the correct second-order relation is 
\bea
    \Delta \rho_2 = \frac{1}{4 \pi G a^2} \big[ \nabla^2 \mathcal{R}_2 + \left( \mathcal{Q} - \nabla^2 \mathcal{C} \right) \big] \,,
\eea
where $\mathcal{Q} \neq \nabla^2 \mathcal{C}$ (see Eq.~(\ref{eq:main:relation})). In the soft limit $k \rightarrow 0$, both $\nabla^2 \mathcal{R}_2$ and $\nabla^2 \mathcal{C}$ vanish while $\mathcal{Q}$ tends towards a finite constant. Thus, the white noise content of $\Delta \rho_2$ resides solely in $\mathcal{Q}$; the term omitted by the linear relation is not a small correction, but rather the entirety of the white noise.

In what follows, we will confirm that the white noise in $\Delta \rho_2$ is genuine, and compute it explicitly for a perfect fluid in radiation domination, including the exact form of the weight function introduced in Ref.~\cite{Noisy}. We will also show, however, that it is \textit{not} inherited by $\mathcal{R}_2$ as an IR enhancement in the way argued in Ref.~\cite{Noisy}. On the contrary, we will find that $\lim_{k \rightarrow 0} (k^2 \mathcal{R}_2) = 0$. The coefficient of the $1/k^2$ pole in $\mathcal{R}_2$ is an exact constant of motion in the soft limit, which cannot be sourced or modified. Because it is initially vanishing, it remains so. Notably, this conservation follows directly from the equation of motion computed from the second-order Einstein equations and holds regardless of the explicit form for the quadratic sources. The upshot is that the purported IR enhancement of $\mathcal{P}_\mathcal{R}(k)$ in Ref.~\cite{Noisy} is an artifact of using a linear expression to relate second-order quantities. The bound on $k_{\rm BH}$, and by extension its implications for the small-scale power spectrum, do not apply.\footnote{We emphasize that this does not render the kurvature white noise itself unphysical. $\Delta \rho$ is a covariant, locally measurable scalar, and the LSWN it develops is a genuine feature of the second-order dynamics, which we compute explicitly below. What fails is the step connecting this white noise to an IR enhancement of the curvature power spectrum.}

The remainder of this paper is structured as follows: In Section~\ref{sec:kurvature:curvature}, we introduce the kurvature density $\Delta \rho$ and establish its relation to the comoving curvature perturbation $\mathcal{R}$ at first and second order. In Section~\ref{sec:RD:kurvature}, we compute $\Delta \rho$ during radiation domination, establishing the explicit form for the quadratic source $\mathcal{Q}$. In Section~\ref{sec:RD:curvature}, we compute $\mathcal{R}$ at second order, and show that the coefficient of the second-order pole necessary for the argument of~\cite{Noisy} is conserved, and moreover vanishing, in the soft limit. We conclude in Section~\ref{sec:discussion} with some discussion of our findings and their implications, as well as identification of several future directions which would be interesting to pursue.

\section{Kurvature and the Curvature Perturbation}\label{sec:kurvature:curvature}

\subsection{The Kurvature Density}

Here we review the ``kurvature density'' $\Delta \rho$ first identified as a useful measure of the curvature inhomogeneity in Refs.~\cite{LSWN,Noisy}, and whose definition was given in Eq.~(\ref{eq:kurvature:def}) as $\Delta \rho = \rho - \theta^2/24 \pi G$, with $\rho = T_{\mu \nu} u^\mu u^\nu$ the energy density and $\theta = \nabla_\mu u^\mu$ the scalar expansion associated with the fluid 4-velocity $u^\mu$. In an unperturbed Friedmann-Lema{\^ i}tre-Robertson-Walker (FLRW) spacetime, for which $\theta = 3 H$, with $H = \dot{a}/a$ the Hubble rate and $a$ the scale factor, the kurvature density is $\Delta \rho = 3 \kappa/8\pi G a^2$. Here, $\kappa$ is the usual curvature constant appearing in the Friedmann equation, which vanishes for a flat cosmology, such that $\Delta \rho = 0$. In a perturbed FLRW spacetime, $\Delta \rho$ will generically be non-zero, even if $\kappa = 0$. Thus, $\Delta \rho$ can be understood as providing a local measure of the deviation from flatness. 

As the sum of two local scalars, $\Delta \rho$ is manifestly local. Its definition is also fully covariant; one can specify $\Delta \rho$ geometrically on the physical spacetime without having to choose coordinates or a perturbative gauge. Indeed, the definition given in Eq.~(\ref{eq:kurvature:def}) is fully non-perturbative, though practically it will be useful to expand it perturbatively in what follows. At linear order, $\Delta \rho_1$ is directly proportional to the local intrinsic spatial curvature, as we will soon see, giving it a particularly nice interpretation. This interpretation will not hold beyond linear order though. These qualities make the kurvature density a useful intermediate quantity for tracking curvature.

An equation of motion for $\Delta \rho$ can be derived by combining the Raychaudhuri equation with the continuity equation. The former reads
\bea\label{eq:Raychaudhuri}
    \dot{\theta} = - \frac{1}{3} \theta^2 - 2\sigma^2 + 2\omega^2 + \nabla_\mu a^\mu - 4\pi G (\rho + 3 p) \,,
\eea
where $\sigma^2 = \frac{1}{2} \sigma_{\mu \nu} \sigma^{\mu \nu}$ is constructed from the trace-free symmetric shear $\sigma_{\mu \nu}$; $\omega^2 = \frac{1}{2} \omega_{\mu \nu} \omega^{\mu \nu}$ is constructed from the antisymmetric vorticity tensor $\omega_{\mu \nu}$; $a^\mu$ is the proper acceleration of $u^\mu$; $p$ is the pressure; and $\dot{X} \equiv u^\mu \nabla_\mu X$ denotes a covariant derivative along the fluid worldlines. See Appendix~\ref{sec:geometry} for definitions of the above quantities and a review of the geometric description of spacelike hypersurfaces. The second piece needed is the continuity equation expressing local energy conservation $u^\mu \nabla_\nu T_\mu^\nu = 0$, or
\bea
    \dot{\rho} + (\rho + p) \theta = \sigma_{\mu \nu} \pi^{\mu \nu} \,,
\eea
where $\pi^{\mu \nu}$ is the anisotropic stress. These equations can then be straightforwardly combined to obtain
\bea\label{eq:kurvature:eom}
    \dot{\Delta \rho} + \frac{2}{3} \theta \Delta \rho = \frac{1}{12 \pi G} \theta \big( 2 \sigma^2 - 2 \omega^2 - \nabla_\mu a^\mu \big) + \sigma_{\mu \nu} \pi^{\mu \nu} \,.
\eea
Notice that the non-linear source $\propto (\sigma^2 - \omega^2)$ carries no spatial derivatives. Such terms will be responsible for the kurvature density developing LSWN. Going forward, we assume vanishing anisotropic stress $\pi_{\mu \nu} = 0$ since it is not necessary for the LSWN effect.

\subsection{Setup and Conventions}

While working in the covariant approach has many advantages, predictions for observable quantities such as the CMB anisotropies are more naturally expressed in terms of metric perturbations. For this reason, we now wish to express the geometric quantities described above in terms of the usual variables of gauge invariant cosmological perturbation theory. We adopt the notation and conventions of Refs.~\cite{Malik:2008im,Christopherson:2011ra}, which we also review in detail in Appendix~\ref{sec:2nd:order:quantities}. In particular, Sections~\ref{sec:metric}, \ref{sec:connection}, and \ref{sec:matter} collect the components of the metric, connection, and fluid 4-velocity. Section~\ref{sec:geometric} expresses the components of the geometric quantities corresponding to the matter 4-velocity $u^\mu$---including the spatial projection tensor, expansion, shear, vorticity, and acceleration---in terms of metric perturbations. Dynamical and constraint equations following from the Einstein field equations and energy-momentum conservation are contained in Appendix~\ref{sec:dynamics:constraints}. Finally, Appendix~\ref{sec:2nd:order:comoving:curvature} reviews the derivation of the comoving curvature perturbation at second order.

Here we simply note that all tensorial quantities are expanded up to second order about a spatially flat FLRW background as
\bea\label{eq:expansion:convention}
    T = T_{0} + \delta T_{1} + \frac{1}{2} \delta T_{2} \,.
\eea
We work in conformal time $\eta$, with $\mathcal{H} = a H$ the conformal Hubble rate and $'$ denoting $\partial_\eta$. The perturbed metric components expressed up to second order read
\bea\label{eq:metric:main}
    g_{00} & = - a^2 \big( 1 + 2 \phi_1 + \phi_2 \big) \,,\\
    g_{0i} & = a^2 \big( B_{1 i} + \frac{1}{2} B_{2 i} \big) \,,\\
    g_{ij} & = a^2 \big( \delta_{ij} + 2 C_{1 ij} + C_{2 ij} \big) \,,
\eea
where at each order $B_i = \partial_i B - S_i$ and
\bea\label{eq:Cij:decomp}
    C_{ij} = - \psi \delta_{ij} + \partial_i \partial_j E + \frac{1}{2} \big( \partial_j F_i + \partial_i F_j \big) + \frac{1}{2} h_{ij} \,,
\eea
with $\partial_i S^i = \partial_i F^i = 0$ and $\partial_j h^{ij} = h_i^i = 0$. At each order, we define the velocity perturbation $v^i$ as the spatial part of the fluid 4-velocity, $u^i = v^i/a$. It decomposes as $v^i = \partial^i v + \mathfrak{v}^i$ with $\partial_i \mathfrak{v}^i = 0$.

At the background level, we make frequent use of the Friedmann and continuity equations
\bea\label{eq:Friedmann:1}
    \mathcal{H}^2 = \frac{8 \pi G}{3} a^2 \rho_0 \,,
\eea
\bea\label{eq:Friedmann:2}
    \mathcal{H}' = - \frac{4 \pi G}{3} a^2 \big(\rho_0 + 3 p_0 \big) \,,
\eea
\bea\label{eq:continuity}
    \rho'_{0} + 3 \mathcal{H} \big(\rho_0 + p_0 \big) = 0 \,.
\eea
Additionally, two first-order results will be used repeatedly below; the Hamiltonian and momentum constraints read 
\bea\label{eq:1st:order:energy:constraint}
    3 \mathcal{H} \big( \psi_1' + \mathcal{H} \phi_1 \big) - \nabla^2 \psi_1 - \mathcal{H} \nabla^2 \big( E_1' - B_1 \big) = - 4 \pi G a^2 \delta \rho_1 \,,
\eea
\bea\label{eq:1st:order:momentum:constraint}
    \psi_1' + \mathcal{H} \phi_1 = - 4 \pi G a^2 \big( \rho_0 + p_0 \big) \big(v_{1} + B_{1} \big) \,.
\eea
The remaining first-order equations, as well as all second-order equations, are collected in Appendix~\ref{sec:dynamics:constraints}.

We work throughout in longitudinal/Poisson gauge, setting $E = B = 0$ at each order. The remaining scalar perturbations are then $\phi$, $\psi$, $v$, $\delta \rho$, and $\delta p$. At first order, the trace-free $ij$ equation gives\footnote{These coincide with the Bardeen potentials $\Phi_1 = \phi_1 - \mathcal{H} \big( E_1' - B_1 \big) - \big( E_1'' - B_1' \big)$ and $\Psi_1 = \psi_1 + \mathcal{H} \big( E_1' - B_1 \big)$, which are gauge-invariant at first order. At second order, one can construct gauge-invariant generalizations $\Phi_2$ and $\Psi_2$, but they will not coincide with $\phi_2$ and $\psi_2$.}
\bea
    \phi_1 = \psi_1 \,.
\eea
Additionally, from the momentum constraint~\eqref{eq:1st:order:momentum:constraint} along with the Friedmann equation~\eqref{eq:Friedmann:1}, we see that the velocity perturbation $v_1$ is fixed in terms of $\psi_1$ and background quantities as
\bea\label{eq:v1:longitudinal:gauge}
    v_1 = - \frac{2}{3(1+w)} \bigg( \frac{\psi_1' + \mathcal{H} \psi_1}{\mathcal{H}^2} \bigg) \,.
\eea

\subsection{Kurvature and Curvature at First Order}

It will be useful to work out the expression for $\Delta \rho$ in terms of the variables of cosmological perturbation theory. Expanding Eq.~(\ref{eq:kurvature:def}) to first order gives
\bea
    \Delta \rho_1 = \delta \rho_1 - \frac{\mathcal{H}}{4 \pi G a} \delta \theta_1 \,.
\eea
Using the expression for $\delta \theta_1$ in Eq.~(\ref{eq:theta:components}) and trading $\delta \rho_1$ for metric perturbations via the Hamiltonian constraint~\eqref{eq:1st:order:energy:constraint}, one can show that the first-order expression is equivalent to the following combination of perturbation variables:
\bea\label{eq:1st:order:kinematic:kurvature}
    \Delta \rho_1 = \frac{1}{4\pi G a^2} \nabla^2 \bigg( \psi_1 - \mathcal{H} \big( v_1 + B_1 \big) \bigg) \,.
\eea
In Poisson gauge, with $E_1 = B_1 = 0$, we have
\bea
    \Delta \rho_1 = \frac{1}{4\pi G a^2} \nabla^2 \big( \psi_1 - \mathcal{H} v_1 \big) \,.
\eea

Before moving on, it is worth noticing that the combination which appears in parentheses in Eq.~(\ref{eq:1st:order:kinematic:kurvature}) is precisely the (first-order) comoving curvature perturbation $\mathcal{R}_1$~\cite{Lyth:1984gv},
\bea\label{eq:comoving:curvature:perturbation}
    \mathcal{R}_1 = \psi_1 - \mathcal{H} \big( v_1 + B_1 \big) \,,
\eea
and so at linear order
\bea\label{eq:kurvature:curvature:1st:order:relation}
    \Delta \rho_1 = \frac{1}{4\pi G a^2} \nabla^2 \mathcal{R}_1 \,.
\eea
The quantity $\mathcal{R}_1$ is gauge invariant, and by Eq.~(\ref{eq:kurvature:curvature:1st:order:relation}) so too will be $\Delta \rho_1$. This is not too surprising; since $\Delta \rho_0$ vanishes at the background level, the first order perturbation was guaranteed to be gauge invariant by the Stewart-Walker lemma. What is interesting is the interpretation of the linear kurvature afforded by its relation to $\mathcal{R}_1$ in Eq.~(\ref{eq:kurvature:curvature:1st:order:relation}); $\Delta \rho_1$ is not the comoving curvature perturbation itself, but rather the curvature density associated with spatial variations in $\mathcal{R}_1$. Further, at linear order the comoving curvature perturbation is simply related to the intrinsic curvature of spatial hypersurfaces ${}^{(3)}\!R_1$ as
\bea
    {}^{(3)}\!R_1 = \frac{4}{a^2} \nabla^2 \mathcal{R}_1 \,.
\eea
Thus, we immediately see that
\bea
    \Delta \rho_1 = \frac{1}{16 \pi G} {}^{(3)}\!R_1 \,.
\eea
At linear order, $\Delta \rho_1$ is effectively the local intrinsic spatial curvature in the units of an energy density (hence the name ``kurvature density''). We emphasize that this will not remain true beyond linear order. 

\subsection{Kurvature and Curvature at Second Order}

Expanding Eq.~(\ref{eq:kurvature:def}) to second order, we find
\bea
    \Delta \rho_2 = \delta \rho_2 - \frac{1}{12 \pi G} \bigg( \frac{3}{a} \mathcal{H} \delta \theta_2 + \delta \theta_1^2 \bigg) \,.
\eea
One can use the Hamiltonian constraint to eliminate $\delta \rho_2$ to find that, schematically, $\Delta \rho_2 \sim {}^{(3)}\!R_2 + \sigma_{(1)}^2 - \omega_{(1)}^2$. Thus, beyond linear order the identification of $\Delta \rho$ with the intrinsic curvature no longer holds; instead, there are also extrinsic contributions from shear and vorticity sourced by quadratic combinations of first-order perturbations. These additional terms contribute to the local deviation from critical expansion and will be important in what follows.

Proceeding analogously to the first-order calculation, one can use the expressions for $\delta \theta_1$ and $\delta \theta_2$ in Eq.~(\ref{eq:theta:components}) and then exchange $\delta \rho_2$ for metric components via the second-order Hamiltonian constraint in Eq.~(\ref{eq:2nd:order:energy:constraint}) to obtain an expression for the second-order kurvature in terms of metric perturbations. In Poisson gauge and restricting to the scalar sector, it reads
\bea\label{eq:delta:rho:2:longitudinal}
    \Delta \rho_2 = \frac{1}{4 \pi G a^2} \bigg[ \nabla^2 \big( \psi_2 - \mathcal{H} v_2 \big) + \mathcal{Q} \bigg] \,,
\eea
where $\mathcal{Q}$ is built from quadratic products of first-order perturbations as
\bea\label{eq:Q:def}
    \mathcal{Q} = \frac{1}{3} \bigg[ & 24 \psi_1 \nabla^2 \psi_1 + 9 \big( \partial_i \psi_1 \big) \big( \partial^i \psi_1 \big) + \frac{8}{(1+w)\mathcal{H}} \big( \partial_i \psi_1 \big) \big( \partial^i X \big) \\
    & - \frac{4(8 + 9w)}{3(1+w)^2 \mathcal{H}^2} \big( \partial_i X \big) \big( \partial^i X \big) - \frac{4}{(1+w) \mathcal{H}^2} X \nabla^2 X \\
    & - \frac{8}{3 (1+w)^2 \mathcal{H}^3} \big( \partial_i X' \big) \big( \partial^i X \big) - \frac{4}{9 (1+w)^2 \mathcal{H}^4} \big( \nabla^2 X \big)^2 \bigg] \,,
\eea
with $w = P_0/\rho_0$ the equation of state and $X \equiv \psi_1' + \mathcal{H} \psi_1$. This object is entirely specified by the Bardeen potential $\psi_1$ and the background evolution.

Comparing Eq.~(\ref{eq:delta:rho:2:longitudinal}) with its first-order counterpart Eq.~(\ref{eq:kurvature:curvature:1st:order:relation}), we note that while at first order the kurvature density is exactly the Laplacian of the comoving curvature perturbation, at second order this is no longer so; while the analogous metric combination $\psi_2 - \mathcal{H} v_2$ appears, this is accompanied by the quadratic sources $\mathcal{Q}$ where the extrinsic curvature lives. Moreover, at second order $\psi_2 - \mathcal{H} v_2$ is not by itself the comoving curvature perturbation. As detailed in Appendix~\ref{sec:2nd:order:comoving:curvature}, we define $\mathcal{R}_2 \equiv \bar{\psi}_2^{\rm co}$ as the value of $\psi_2$ in a gauge with comoving slicing $\bar{v}_1 + \bar{B}_1 = 0$ and isotropic threading $\bar{E}_1 = 0$, such that $\psi_2$ measures the intrinsic geometry of the slice. The result is
\bea\label{eq:R2:main:text}
    \mathcal{R}_2 = \psi_2 - \mathcal{H} v_2 + \mathcal{C} \,,
\eea 
with the quadratic completion given by
\bea\label{eq:C:main:text}
    \mathcal{C} = \frac{2}{3(1+w)} \bigg[ & \frac{2}{3(1+w) \mathcal{H}^3} \nabla^{-2} \partial^i \big( X' \partial_i X - X \partial_i X'\big) - \frac{6}{\mathcal{H}} \nabla^{-2} \partial^i \left( \psi_1 \partial_i X \right) \\
    & - \frac{2}{\mathcal{H}} \psi_1 X + \frac{2}{3(1+w)} \frac{\mathcal{H}'}{\mathcal{H}^4} X^2 - \frac{2(5+3w)}{3(1+w)} \frac{1}{\mathcal{H}^2} X^2 - \frac{2}{3(1+w) \mathcal{H}^3} X X' \\
    & + \frac{1}{3(1+w) \mathcal{H}^4} (\partial_i X) (\partial^i X) - \frac{1}{3(1+w) \mathcal{H}^4} \nabla^{-2} \partial_i \partial_j \left( \partial^i X \partial^j X \right) \bigg] \,.
\eea
Substituting Eq.~(\ref{eq:R2:main:text}) into Eq.~(\ref{eq:delta:rho:2:longitudinal}), one finds the exact second order relation 
\bea\label{eq:main:relation}
    \Delta \rho_2 = \frac{1}{4 \pi G a^2} \bigg[ \nabla^2 \mathcal{R}_2 + \left( \mathcal{Q} - \nabla^2 \mathcal{C} \right) \bigg] \,.
\eea

Meanwhile, the relation posited in Ref.~\cite{Noisy} to convert the white noise in the kurvature density to the comoving curvature perturbation reads
\bea\label{eq:Noisy:relation}
    \nabla^2 \mathcal{R} \stackrel{?}{=} 4 \pi G a^2 \Delta \rho \,.
\eea
In order for this to be true, one would require that $\mathcal{Q} \stackrel{?}{=} \nabla^2 \mathcal{C}$. Already from Eqs.~(\ref{eq:Q:def}) and (\ref{eq:C:main:text}), however, we see that these are very different objects by construction. There is no reason why they should coincide, and as we will see explicitly in the coming sections, they do not in general. 

In the remainder of this paper, we will establish the two quadratic terms of Eq.~(\ref{eq:main:relation}) separately. In Sec.~\ref{sec:RD:kurvature} we compute $\mathcal{Q}$ explicitly during radiation domination and show that it is non-vanishing in the limit $k \rightarrow 0$, implying that the white noise in $\Delta \rho_2$ derived in Ref.~\cite{Noisy} is genuine. In Sec.~\ref{sec:RD:curvature}, we compute $\mathcal{C}$ explicitly during radiation domination and show that $\mathcal{Q} \neq \nabla^2 \mathcal{C}$ and $\lim_{k \rightarrow 0} (k^2 \mathcal{C}) = 0$. Further, we will show that $\lim_{k \rightarrow 0} (k^2 \mathcal{R}_2) = 0$. The implication is that the white noise in $\Delta \rho_2$ does \textit{not} imply a pole in $\mathcal{R}_2$; rather, it is carried entirely by $\mathcal{Q}$. The bound in Ref.~\cite{Noisy} therefore does not follow.

\section{Kurvature during Radiation Domination}\label{sec:RD:kurvature}

In order to evaluate the quadratic source $\mathcal{Q}$ in Eq.~(\ref{eq:Q:def}), we simply need to solve for $\psi_1$. In Poisson gauge, Eq.~(\ref{eq:1st:order:psi:eom}) simplifies to 
\bea\label{eq:psi1:eom:general}
    \psi_1'' + 3 \mathcal{H} \big( 1 + c_s^2 \big) \psi_1' + \big( 2 \mathcal{H}' + (1 + 3 c_s^2) \mathcal{H}^2 - c_s^2 \nabla^2 \big) \psi_1= 0 \,,
\eea
where $c_s^2 \equiv \delta p_1/\delta \rho_1$ is the adiabatic sound speed and we have also used~\eqref{eq:1st:order:energy:constraint}. For arbitrary equation of state, the scale factor and conformal Hubble rate appearing in this equation are
\bea\label{eq:a:Hubble:arbitrary:EOS}
    a(\eta) = a_{\rm ref} \bigg( \frac{\eta}{\eta_{\rm ref}} \bigg)^{2/(1+3w)} \,, \quad \mathcal{H}(\eta) = \frac{2}{(1+3w) \eta} \,.
\eea

Let us now specialize to radiation domination, for which $w = c_s^2 = 1/3$ and so $a \propto \eta$ and $\mathcal{H} = 1/\eta$. Let us also Fourier transform, which is trivial at the linear level. The equation for the Fourier transformed $\tilde{\psi}_1(\eta,\vec{q}\,)$ becomes
\bea\label{eq:psi1:eom:RD}
    \tilde{\psi}_1'' + 4 \mathcal{H} \tilde{\psi}_1' + \frac{1}{3} q^2 \tilde{\psi}_1 = 0 \,.
\eea
It is convenient to decompose the field into a primordial perturbation and the deterministic transfer function, $\tilde{\psi}_1(\eta, \vec{q}\,) \equiv T_\psi(\eta,q) \tilde{\psi}_1^{\rm prim}(\vec{q}\,)$, where $\lim_{\eta \rightarrow 0} T_\psi(\eta,q) = 1$. The transfer function satisfying Eq.~(\ref{eq:psi1:eom:RD}) is explicitly
\bea
    T_\psi(\eta,q) = 3 \bigg( \frac{\sin \big(q\eta/\sqrt{3} \big)}{\big(q\eta/\sqrt{3}\big)^3} - \frac{\cos\big(q\eta/\sqrt{3}\big)}{\big(q\eta/\sqrt{3}\big)^2} \bigg) \,.
\eea
We can further relate $\tilde{\psi}_1^{\rm prim}(\vec{q}\,)$ to the primordial curvature perturbation $\mathcal{R}_1^{\rm prim}(\vec{q}\,)$ as $\mathcal{R}_1^{\rm prim} = \frac{5+3w}{3(1+w)} \psi_1^{\rm prim}$, such that
\bea
    \tilde{\psi}_1(\eta, \vec{q}\,) = \frac{3(1+w)}{5+3w} T_\psi(\eta,q) \mathcal{R}_1^{\rm prim}(\vec{q}\,) \,.
\eea
Similarly, we decompose the Fourier transform of $X = \psi_1' + \mathcal{H} \psi_1$ as
\bea
    \tilde{X}(\eta,\vec{q}\,) = \frac{3(1+w)}{5+3w} T_X(\eta,q) \mathcal{R}_1^{\rm prim}(\vec{q}\,) \,,
\eea
with $T_X \equiv T_\psi' + \frac{1}{\eta} T_\psi$. We will drop the ``prim'' superscript on $\mathcal{R}_1$ going forward, leaving it implicit that this is the primordial quantity.

Using these decompositions, we may write the Fourier transform of Eq.~(\ref{eq:delta:rho:2:longitudinal}) during radiation domination as
\bea\label{eq:Delta:rho:2:FT}
    \widetilde{\Delta \rho}_2(\eta,\vec{k}) = - \frac{k^2}{4 \pi G a^2} \bigg( \tilde{\psi}_2(\eta,\vec{k}) - \frac{1}{\eta} \tilde{v}_2(\eta,\vec{k}) \bigg) + \int \frac{\dd^3 q}{(2\pi)^3} f\big(\eta,\vec{k},\vec{q}\,\big) \mathcal{R}_1(\vec{q}\,) \mathcal{R}_1(\vec{k} - \vec{q}\,) \,,
\eea
where we have defined the kernel
\bea
    f\big(\eta,\vec{k},\vec{q}\,\big) = \frac{1}{108\pi G a^2} \bigg[ & 3 \eta^3 \vec{q} \cdot (\vec{k} - \vec{q}\,) \bigg( T_X(\eta,q)' T_X(\eta,|\vec{k} - \vec{q}\,|) + T_X(\eta,q) T_X(\eta,|\vec{k} - \vec{q}\,|)' \bigg) \\
    & + \bigg( 6 \eta^2 (q^2 + |\vec{k} - \vec{q}\,|^2 ) + 33 \eta^2 \vec{q} \cdot (\vec{k} - \vec{q}\,) + \eta^4 q^2 |\vec{k} - \vec{q}\,|^2 \bigg) T_X(\eta,q) T_X(\eta,|\vec{k} - \vec{q}\,|) \\
    & - 12 \eta \vec{q} \cdot (\vec{k} - \vec{q}\,) \bigg( T_X(\eta,q) T_\psi(\eta,|\vec{k} - \vec{q}\,|) + T_\psi(\eta,q) T_X(\eta,|\vec{k} - \vec{q}\,|) \bigg) \\
    & - 12 \bigg( 4 (q^2 + |\vec{k} - \vec{q}\,|^2) + 3 \vec{q} \cdot (\vec{k} - \vec{q}\,) \bigg) T_\psi(\eta,q) T_\psi(\eta,|\vec{k} - \vec{q}\,|) \bigg] \,,
\eea
which is the Fourier-space representation of the quadratic source $\mathcal{Q}$ defined in Eq.~(\ref{eq:Q:def}) evaluated on the radiation domination transfer functions. Note also that we have symmetrized this kernel in anticipation of simplifying the computation of the power spectrum, which is defined in the usual way
\bea\label{eq:Pk:def}
    \langle \widetilde{\Delta \rho}_2(\vec{k}) \widetilde{\Delta \rho}_2(\vec{k}') \rangle = (2\pi)^3 \delta^{(3)}(\vec{k} + \vec{k}') P_{\Delta \rho_2}(k) \,.
\eea
Substituting Eq.~(\ref{eq:Delta:rho:2:FT}) into the left-hand side of the above, we find 
\bea\label{eq:Delta:rho:2:correlator}
    \langle \widetilde{\Delta \rho}_2(\vec{k}) \widetilde{\Delta \rho}_2(\vec{k}') \rangle = & \frac{k^2 {k'}^2}{16 \pi^2 G^2 a^4} \left\langle \left( \tilde{\psi}_2(\vec{k}) - \mathcal{H} \tilde{v}_2(\vec{k}) \right) \left( \tilde{\psi}_2(\vec{k}') - \mathcal{H} \tilde{v}_2(\vec{k}') \right) \right\rangle + [\text{cross terms}] \\
    & + \int \!\frac{\dd^3 q}{(2\pi)^3} \int \!\frac{\dd^3 q'}{(2\pi)^3} f(\eta,\vec{k},\vec{q}\,) f(\eta,\vec{k}',\vec{q}\,') \big\langle \mathcal{R}_1(\vec{q}\,) \mathcal{R}_1(\vec{k} - \vec{q}\,) \mathcal{R}_1(\vec{q}\,') \mathcal{R}_1(\vec{k}' - \vec{q}\,') \big\rangle \,.
\eea
The cross terms involve the three-point function of the primordial $\mathcal{R}_1$ with second order quantities, which vanish for Gaussian initial conditions, and so we drop them without loss of generality. Further, under the Gaussian statistics, the 4-point function in the second line factorizes as
\bea\label{eq:Wick}
    \big\langle \mathcal{R}_1(\vec{q}\,) \mathcal{R}_1(\vec{k} - \vec{q}\,) \mathcal{R}_1(\vec{q}\,') \mathcal{R}_1(\vec{k}' - \vec{q}\,') \big\rangle = (2\pi)^6 \big( \delta^{(3)}(\vec{q} + \vec{q}\,') + \delta^{(3)}(\vec{q} + \vec{k}' - \vec{q}\,') \big) \delta^{(3)}(\vec{k} + \vec{k}') P_{\mathcal{R}_1}(q) P_{\mathcal{R}_1}(|\vec{k} - \vec{q}\,|) \,.
\eea
Note that both terms give the same contribution because we have symmetrized the kernel $f$.

Since we are ultimately interested in white noise on large scales, let us now take the limit $k \rightarrow 0$. The term in the first line of~\eqref{eq:Delta:rho:2:correlator} naively appears to scale as $k^4$, and so should vanish in the soft limit. This is only true provided the combination $\tilde{\psi}_2 - \mathcal{H} \tilde{v}_2$ is finite in this limit. In the next section, we will prove rigorously that this is the case by solving the second order Einstein equations for $\tilde{\psi}_2$ and $\tilde{v}_2$. We will see that $\lim_{k \rightarrow 0} k^2(\tilde{\psi}_2 - \mathcal{H} \tilde{v}_2)$ is a constant of motion, and moreover, that because its initial value is vanishing, it remains so: $\lim_{k \rightarrow 0} k^2(\tilde{\psi}_2 - \mathcal{H} \tilde{v}_2) = 0$. (See Eq.~(\ref{eq:lim:psi2:v2}) and the preceding discussion.) The first line thus vanishes in the limit $k \rightarrow 0$, and all that remains is the second line. 

Meanwhile, in the soft limit, the kernel reduces to $f(\eta,\vec{k},\vec{q}) \rightarrow f(\eta,q)$, with 
\bea
    f(\eta,q) = \frac{1}{108 \pi G a^2} \bigg( \big( q^4 \eta^4 - 21 q^2 \eta^2 \big) T_X(q)^2 - 6 q^2 \eta^3 T_X'(q) T_X(q) + 24 q^2 \eta T_X(q) T_\psi(q) - 60 q^2 T_\psi(q)^2 \bigg) \,.
\eea
Substituting this into Eq.~(\ref{eq:Delta:rho:2:correlator}), using the delta functions in the decomposition~\eqref{eq:Wick} to eliminate the second integral, and comparing with Eq.~(\ref{eq:Pk:def}), we see that
\bea\label{eq:P:LSWN}
    P_{\Delta \rho_2}^{\rm LSWN} \equiv \lim_{k \rightarrow 0} P_{\Delta \rho_2}(k) = 4\pi^2 \int_0^\infty \frac{\dd q}{q^4} f(\eta,q)^2 \mathcal{P}_{\mathcal{R}_1}(q)^2 \,,
\eea
where we have expressed the right-hand side in terms of dimensionless power spectra $\mathcal{P}_{\mathcal{R}_1}(q) = \frac{q^3}{2\pi^2} P_{\mathcal{R}_1}(q)$. The result is a constant, scale-independent (white noise) contribution to the large scale kurvature power spectrum, consistent with the claims of Refs.~\cite{LSWN,Noisy}, now expressed explicitly in terms of radiation domination first-order quantities. So LSWN in the kurvature density field is genuine. However, this does not translate to an IR enhancement of the curvature perturbation, as we will now show.

Ref.~\cite{Noisy} goes on to define the quantity $k_{\rm BH}$, which in our notation reads\footnote{Note that the numerical prefactor differs from that of Refs.~\cite{LSWN,Noisy} since we use a different Fourier convention.}
\bea\label{eq:kBH:def}
    k_{\rm BH} = 8 G^2 a^4 P_{\Delta \rho_2}^{\rm LSWN} \,.
\eea
This object comes from relating the power spectrum for the kurvature density with that of the comoving curvature perturbation, using the linear relation in Eq.~(\ref{eq:Noisy:relation}) to write\footnote{Note that Ref.~\cite{Noisy} denotes the dimensionless power spectrum $\mathcal{P}_\mathcal{R}$ as $\Delta_{\mathcal{R}}^2$.}
\bea
    P_{\rm \Delta \rho}(k) \stackrel{?}{=} \frac{k^4}{(4\pi G a^2)^2} P_{\mathcal{R}}(k) = \frac{k}{8 G^2 a^4} \mathcal{P}_{\mathcal{R}}(k) \quad \rightarrow \quad \mathcal{P}_{\mathcal{R}}(k) \stackrel{?}{=} 8 G^2 a^4 \frac{P_{\Delta\rho}(k)}{k} \,.
\eea
This is the basis on which Ref.~\cite{Noisy} argued that white noise in $\Delta \rho$ (i.e. $\lim_{k \rightarrow 0} P_{\Delta \rho}(k) = \text{constant}$) should give a relic contribution $\propto 1/k$ to the power spectrum of $\mathcal{R}$. There, the total power spectrum was approximated as
\bea\label{eq:bad:parameterization}
    \mathcal{P}_{\mathcal{R}}(k) \stackrel{?}{=} \mathcal{P}_{\mathcal{R}_1}(k) + \frac{k_{\rm BH}}{k} \,,
\eea
with $k_{\rm BH}$ defined in Eq.~\eqref{eq:kBH:def} the scale at which the relic contribution becomes order 1. The bound on $k_{\rm BH}$ was derived by demanding that the contribution from this term not exceed the observed power in CMB anisotropies at small angular multipoles. 

Note, however, that the apparent IR divergence in Eq.~(\ref{eq:bad:parameterization}) is an artifact of using the first-order dictionary to relate first- and second-order quantities. This is inconsistent. In reality, the second-order $\mathcal{R}_2$ and $\Delta \rho_2$ are related by Eq.~(\ref{eq:main:relation}), which features an additional piece $(\mathcal{Q} - \nabla^2 \mathcal{C})$. As we will see in Sec.~\ref{sec:RD:curvature}, $\nabla^2 \mathcal{C}$ vanishes in the limit $k \rightarrow 0$, and so the entire white noise content of $\Delta \rho_2$ is carried by $\mathcal{Q}$. This is precisely the term that Eq.~(\ref{eq:Noisy:relation}) omits, and is the origin of the inconsistency.

\begin{figure}[t!]
\centering
\includegraphics[width=0.45\textwidth]{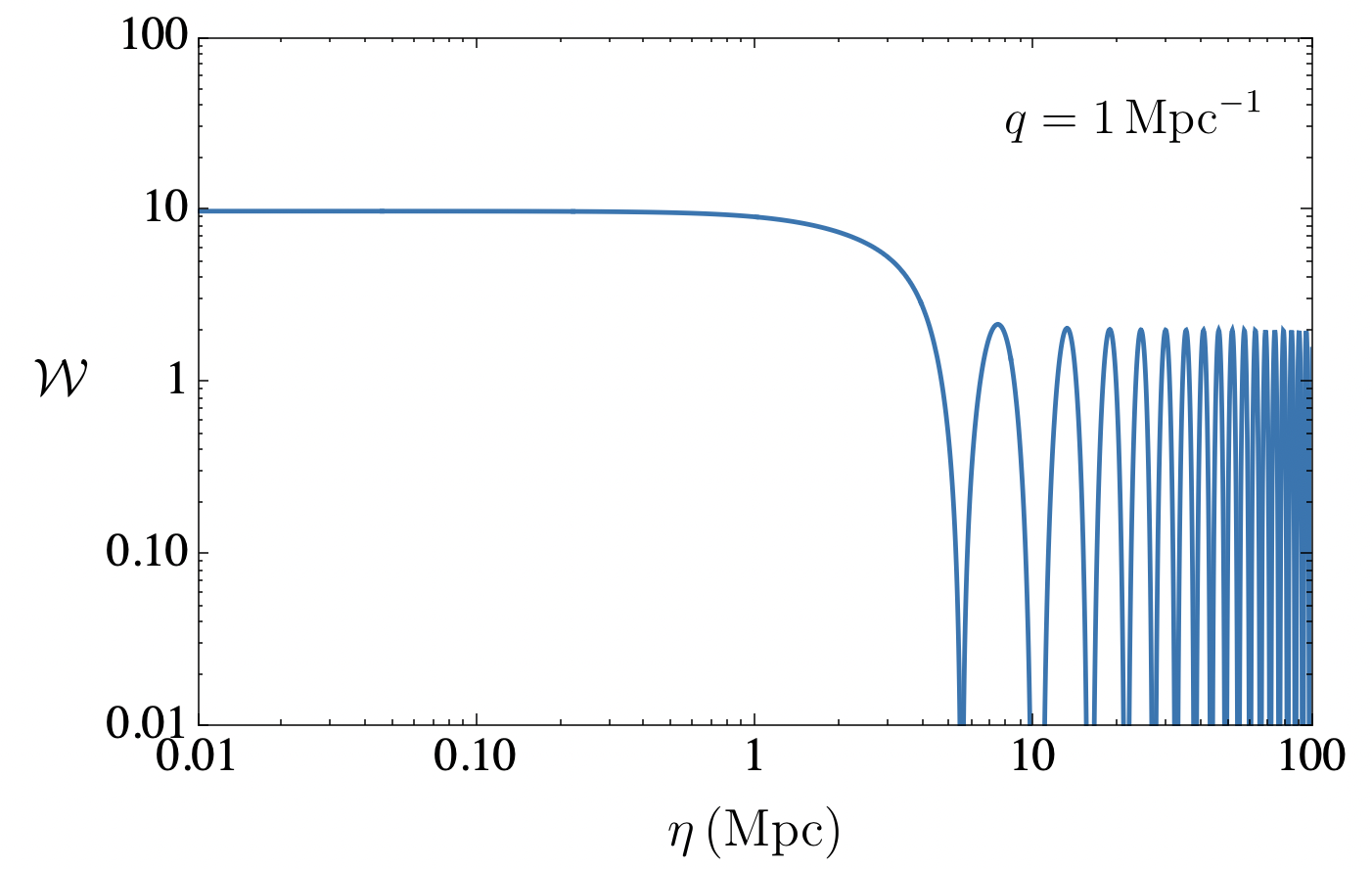}
\hspace{5mm}
\includegraphics[width=0.45\textwidth]{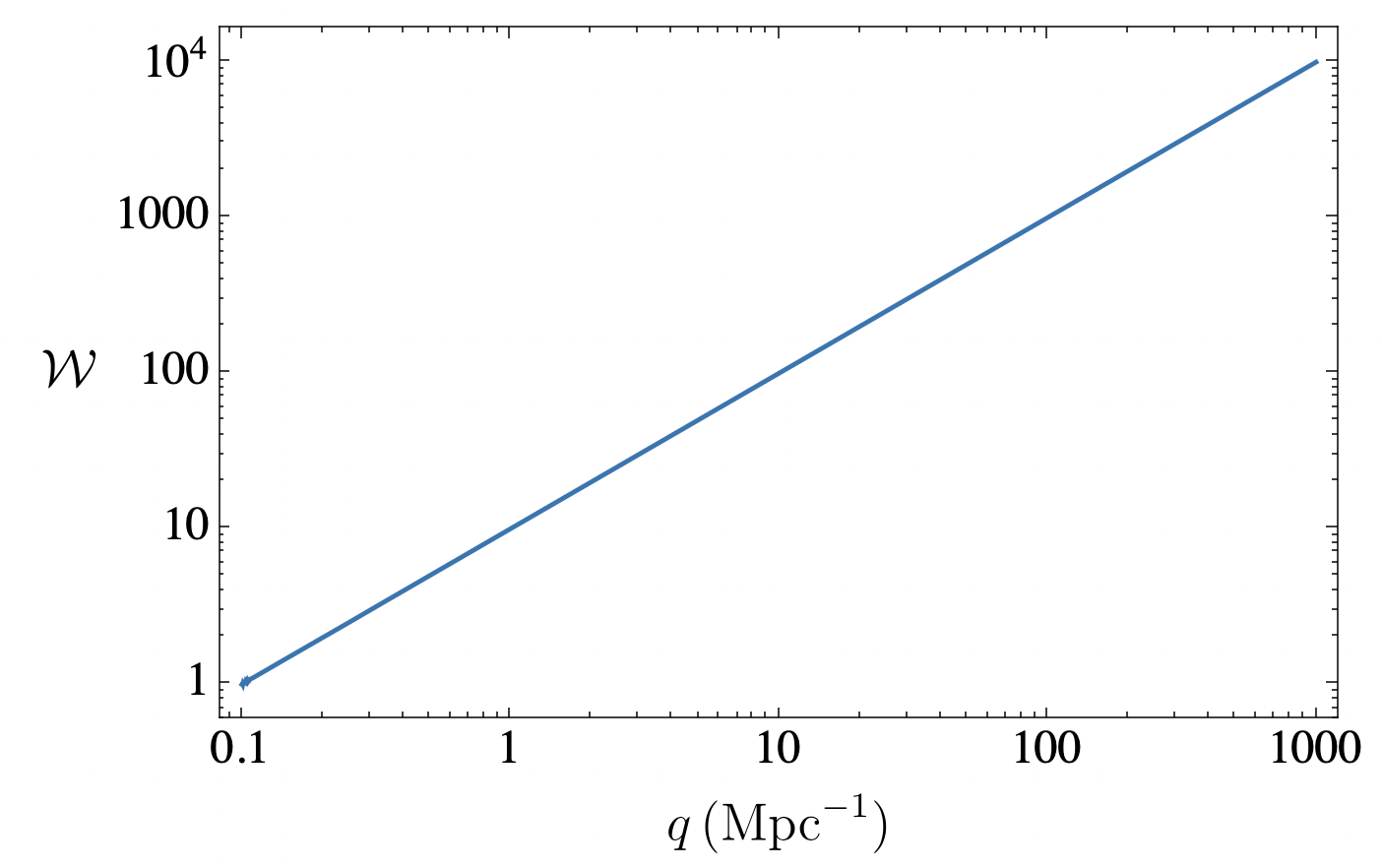}
\caption{\textbf{Left:} Behavior of the kernel~\eqref{eq:W:kernel:def} at fixed wavenumber for evolving conformal time. Notice that the value is robust to the time at which it is evaluated, provided this is done while the mode is still super-horizon ($q\eta < 1$). \textbf{Right:} Behavior of~\eqref{eq:W:kernel:def} as a function of wavenumber, evaluated for each mode at super-horizon times. The growth with $q$ implies that the integral should be dominated by small scales for a scale-invariant $\mathcal{P}_{\mathcal{R}_1}$.}
\label{fig:W}
\end{figure}

While the bound on $k_{\rm BH}$ is not applicable, it is nevertheless interesting to examine the white noise spectrum of $\Delta \rho_2$ in more detail, given that this quantity should in principle be measurable. Using Eq.~(\ref{eq:P:LSWN}), we rewrite $k_{\rm BH}$ as
\bea\label{eq:W:kernel:def}
    k_{\rm BH} = \int_{-\infty}^\infty \dd \ln q \, \mathcal{W}(\eta,q) \mathcal{P}_{\mathcal{R}_1}(q)^2 \,, \quad\quad \mathcal{W}(\eta,q) = 32 \pi^2 G^2 a^4 \frac{f(\eta,q)^2}{q^3} \,,
\eea
where we have defined the kernel $\mathcal{W}$ to be the analog of the ``weight function'' in Eq.~(2) of~\cite{Noisy}. We plot the behavior of $\mathcal{W}$ in Fig.~\ref{fig:W}.
First, notice from the left panel that at fixed comoving wavenumber, the integrand asymptotes to a constant at early times while the mode is still super-horizon. This is consistent with the expected behavior argued for in~\cite{LSWN,Noisy}, and indicates that $k_{\rm BH}$ should be independent of the evaluation time (provided this is while the mode is still super-horizon). Once the mode re-enters, the amplitude decays mildly before oscillating with increasing frequency and a non-decaying amplitude. From the right panel, we see that $\mathcal{W}$ grows with $q$, which was the basis for Ref.~\cite{Noisy}'s argument that the integral should be bounded by having some feature to cut off the small-scale power of $\mathcal{P}_{\mathcal{R}_1}$.

Before moving on, let us summarize the results of this section. The white noise in the kurvature density $\Delta \rho_2$ is genuine, and is encoded in $\mathcal{Q}$, which is finite and non-vanishing as $k \rightarrow 0$. However, what does not follow is the inference to a constraint on $\mathcal{P}_{\mathcal{R}_1}$; this step in~\cite{Noisy} required the linear relation of Eq.~(\ref{eq:Noisy:relation}), or equivalently a cancellation between the quadratic terms in Eq.~(\ref{eq:main:relation}). In the next section, we will compute $\mathcal{C}$ directly and show that $\nabla^2 \mathcal{C} \neq \mathcal{Q}$. Further, we will show that $\lim_{k \rightarrow 0} \left(k^2 \mathcal{R}_2 \right) = 0$, and so there is no IR enhancement of the curvature power spectrum at second order in the manner suggested by Eq.~(\ref{eq:bad:parameterization}).

\section{Curvature during Radiation Domination}\label{sec:RD:curvature}

\subsection{Second-Order System}


From Appendix~\ref{sec:dynamics:constraints}, the second-order Einstein equations and energy-momentum conservation equations in Poisson gauge give the following system:
\begin{subequations}\label{eq:EE:EM}
\bea\label{eq:EE1}
    \nabla^2 \psi_2 - 3 \mathcal{H} (\psi_2' + \mathcal{H} \phi_2) + Q^{(1)} = 4 \pi G a^2 \delta \rho_2 \,,
\eea
\bea\label{eq:EE2}
    \partial_i (\psi_2' + \mathcal{H} \phi_2) + Q_i^{(2)} = - 4\pi G a^2 \rho_0 (1+w) \partial_i v_2 \,,
\eea
\bea\label{eq:EE3}
    3 \psi_2'' + 3 \mathcal{H} (2 \psi_2' + \phi_2') + \nabla^2 (\phi_2 - \psi_2) + 3 \left( 2 \frac{a''}{a} - \mathcal{H}^2 \right) \phi_2 + Q^{(3)}= 12 \pi G a^2 c_{\rm eff,2}^2 \delta \rho_2 \,,
\eea
\bea\label{eq:EM1}
    \delta \rho_2' + 3 \mathcal{H} (1 + c_{\rm eff,2}^2) \delta\rho_2 + (1+w)\rho_0 (\nabla^2 v_2 - 3 \psi_2') + Q^{(4)} = 0 \,,
\eea
\bea\label{eq:EM2}
    (1+w) \rho_0 \left( \frac{\rho_0'}{\rho_0} \partial_i v_2 + \partial_i v_2' + \partial_i \phi_2 + 4 \mathcal{H} \partial_i v_2\right) + c_{\rm eff,2}^2 \partial_i \delta\rho_2 + Q^{(5)}_i = 0 \,,
\eea
\end{subequations}
where $c_{\rm eff,2}^2 \equiv \delta P_2/\delta \rho_2$, and the quadratic first-order sources $Q^{(1)}, Q^{(2)}_i, Q^{(3)}, Q^{(4)}, Q^{(5)}_i$ are collected in Appendix~\ref{sec:quad:sources}. Note that the Bianchi identity relates the $00$, $0i$, $ij$ trace, and energy conservation equations, and so of these 5 equations, only 4 are independent. Having already solved for the first-order quantities, we thus have 4 equations and 4 unknowns: $\psi_2$, $\phi_2$, $v_2$, and $\delta \rho_2$. 

As in Sec.~\ref{sec:RD:kurvature}, we consider radiation domination, for which $\rho = 3 P$ holds non-perturbatively and so $w = c_s^2 = c_{\rm eff,2}^2 = 1/3$. Additionally from Eq.~(\ref{eq:a:Hubble:arbitrary:EOS}), $a \propto \eta$ and $\mathcal{H} = 1/\eta$. Using these relations as well as the Friedmann~\eqref{eq:Friedmann:1} and continuity~\eqref{eq:continuity} equations, one can combine~\eqref{eq:EE1} and~\eqref{eq:EE3} to obtain
\bea\label{eq:Aeq}
    3 \psi_2'' + 9 \mathcal{H} \psi_2' + 3 \mathcal{H} \phi_2' + \nabla^2 \phi_2 - 2 \nabla^2 \psi_2 = Q^{(1)} - Q^{(3)} \,,
\eea
as well as Eqs.~\eqref{eq:EE1},~\eqref{eq:EE2}, and~\eqref{eq:EM2} to obtain
\bea\label{eq:Beq}
    3 \psi_2'' + 9 \mathcal{H} \psi_2' + 3 \mathcal{H} \phi_2' - \nabla^2 \psi_2 = \Sigma \,,
\eea
where we have defined the effective combinations of sources
\bea\label{eq:Sigma:def}
    \Sigma = Q^{(1)} - 6 \mathcal{H} q^{(2)} - 3 {q^{(2)}}' + 6 \mathcal{H}^2 q^{(5)} \,, 
\eea
with
\bea
    q^{(2)} \equiv \nabla^{-2} \partial^i Q_i^{(2)} \,, \,\,\, q^{(5)} \equiv \frac{3}{4\rho_0} \nabla^{-2} \partial^i Q_i^{(5)} \,.
\eea
Additionally, combining~\eqref{eq:Aeq} and~\eqref{eq:Beq} gives
\bea\label{eq:2nd:order:slip}
    \nabla^2 (\psi_2 - \phi_2) = \Xi \,,
\eea
with 
\bea\label{eq:Xi:def}
    \Xi = Q^{(3)} - 6 \mathcal{H} q^{(2)} - 3 {q^{(2)}}' + 6 \mathcal{H}^2 q^{(5)} \,,
\eea
Finally we arrive at the equation of motion for $\psi_2$, 
\bea
    \psi_2'' + 4 \mathcal{H} \psi_2' - \frac{1}{3} \nabla^2 \psi_2 = \mathcal{S} \,, \quad \mathcal{S} \equiv \frac{1}{3} \Sigma + \mathcal{H} \nabla^{-2} \Xi' \,,
\eea
or in Fourier space
\bea\label{eq:2nd:order:psi2:eom}
    \tilde{\psi}_2'' + 4 \mathcal{H} \tilde{\psi}_2' + \frac{1}{3} k^2 \tilde{\psi}_2 = \tilde{\mathcal{S}}_k \,, \quad \tilde{\mathcal{S}}_k \equiv \frac{1}{3} \tilde{\Sigma}_k - \frac{\mathcal{H}}{k^2} \tilde{\Xi}_k' \,.
\eea
The homogeneous operator coincides with Eq.~(\ref{eq:psi1:eom:RD}), but the right-hand side is now sourced.

\subsection{Soft Behavior of Sources}

Let us consider now the curvature perturbation at second order,
\bea\label{eq:R2:Poisson}
    \mathcal{R}_2 = \psi_2 - \mathcal{H} v_2 + \mathcal{C} \,, 
\eea
where $\mathcal{C}$ are products of first order perturbations needed to make $\mathcal{R}_2$ gauge invariant. We will compute what these are momentarily; let us first check that the second order terms themselves do not have any poles. More concretely, the claim in~\cite{Noisy} using the first order Poisson equation was that $\lim_{k \rightarrow 0} \left( k^2 \mathcal{R}_2 \right) = \text{finite}$ due to the white noise in $\Delta \rho_2$. We seek to show that this was an artifact of mixing first and second order equations in an erroneous way; instead we will show that $\lim_{k \rightarrow 0} \left( k^2 \mathcal{R}_2 \right) = 0$ despite the white noise in $\Delta \rho_2$. 

To this end, consider first $\tilde{\psi}_2$ obtained by solving Eq.~(\ref{eq:2nd:order:psi2:eom}) and examine its leading order behavior in $\epsilon \equiv k/q$, with $q$ an internal loop momentum. Note that the scalar sources $Q^{(1)}$ and $Q^{(3)}$ are manifestly local, and so are trivially $\mathcal{O}(\epsilon^0)$. The sources $q^{(2)} \sim \nabla^{-2} \partial^i Q_i^{(2)}$ and $q^{(5)} \sim \nabla^{-2} \partial^i Q_i^{(5)}$ naively appear $\sim \epsilon^{-1}$. Note, however, that what appears in $Q_i^{(2)}$ and $Q_i^{(5)}$ are terms of the form $A \partial_i B$, which upon Fourier transforming in a symmetrized way gives
\bea\label{eq:safe:pole}
    \nabla^{-2} \partial^i \left( A \partial_i B \right) & \stackrel{\text{FT}}{\longrightarrow} \frac{1}{2} A_{\vec{q}} B_{\vec{k} - \vec{q}} + \frac{\mu}{2\epsilon} \left( A_{\vec{k} - \vec{q}} B_{\vec{q}} - A_{\vec{q}} B_{\vec{k} - \vec{q}} \right) \\
    & \stackrel{k \rightarrow 0}{\longrightarrow} \frac{1}{2} A_{q} B_{q} + \frac{\mu^2 q}{2} ( A \partial_q B - B \partial_q A) + \mathcal{O}(\epsilon) \,,
\eea
where we have defined $\mu = \hat{k} \cdot \hat{q}$ and, in passing to the second line, used $|\vec{k} - \vec{q}\,| = q \left( 1 - \epsilon \mu \right) + \mathcal{O}(\epsilon^2)$. Notice that the apparent pole multiplies the combination antisymmetric under exchange of the two hard legs, which is itself $\mathcal{O}(\epsilon)$ and so cancels against the $1/\epsilon$ pole to give something finite in the limit $\vec{k} \rightarrow 0$. Thus, $q^{(2)}$ and $q^{(5)}$ are also $\mathcal{O}(\epsilon^0)$. 

From the definitions in Eqs.~\eqref{eq:Sigma:def} and~\eqref{eq:Xi:def}, both scalar sources are then pole-free, 
\bea
    \tilde{\Sigma} = \mathcal{O}(\epsilon^0) \,, \quad \tilde{\Xi} = \mathcal{O}(\epsilon^0) \,,
\eea
and the only potential pole comes from the explicit $1/k^2$ in the second term in the source $\tilde{\mathcal{S}}_k$ in~\eqref{eq:2nd:order:psi2:eom}. Let us then write $\tilde{\psi}_2$ as $\tilde{\psi}_2 \equiv P/k^2 + \mathcal{O}(\epsilon^0)$, with $P$ the coefficient of the pole, and also isolate the $\mathcal{O}(\epsilon^0)$ part of the source by writing $\tilde{\Xi} = \tilde{\Xi}_0 + \mathcal{O}(\epsilon)$. At $\mathcal{O}(\epsilon^{-2})$, Eq.~(\ref{eq:2nd:order:psi2:eom}) reads
\bea\label{eq:P:eom}
    P'' + 4 \mathcal{H} P' = - \mathcal{H} \tilde{\Xi}_0' \,.
\eea
Now, recall from Eq.~(\ref{eq:R2:Poisson}) that the combination that matters is $\psi_2 - \mathcal{H} v_2$. From Eq.~(\ref{eq:EE2}), we have
\bea
    \tilde{\psi}_2 - \mathcal{H} \tilde{v}_2 = \frac{3}{2} \tilde{\psi}_2 + \frac{\tilde{\psi}_2'}{2\mathcal{H}} + \frac{\tilde{q}^{(2)}}{2\mathcal{H}} + \frac{\tilde{\Xi}}{2k^2} \,,
\eea
where we have also used Eq.~(\ref{eq:2nd:order:slip}) to substitute $\tilde{\phi}_2 = \tilde{\psi}_2 + \tilde{\Xi}/k^2$. As already established, $q^{(2)}$ is pole-free, so the $\mathcal{O}(\epsilon^{-2})$ piece vanishes provided
\bea
    P' + 3 \mathcal{H} P = - \mathcal{H} \tilde{\Xi}_0 \,.
\eea
Notice that this combination is a first integral of the equation of motion; upon differentiating one recovers Eq.~(\ref{eq:P:eom}). So this is not actually an extra condition on the dynamics, but rather implied by them, potentially up to an integration constant. The conclusion is that the coefficient of the $\mathcal{O}(\epsilon^{-2})$ term in $\mathcal{R}_2$ is a constant of motion in the soft limit, and therefore cannot be sourced. 

\subsection{Conserved Pole Coefficient}

Let us make this a bit more precise. We define the combination
\bea\label{eq:F:def}
    \mathcal{F} \equiv \frac{P'}{\mathcal{H}} + 3 P + \tilde{\Xi}_0 \,,
\eea
such that
\bea\label{eq:lim:is:F}
    \lim_{k \rightarrow 0} k^2 (\tilde{\psi}_2 - \mathcal{H} \tilde{v}_2 ) = \frac{1}{2} \mathcal{F} \,.
\eea
Differentiating Eq.~(\ref{eq:F:def}) and using the equation of motion~\eqref{eq:P:eom}, one obtains
\bea
    \mathcal{F}' = \frac{1}{\mathcal{H}} \left( P'' + 4 \mathcal{H} P' + \mathcal{H} \tilde{\Xi}_0' \right) = 0 \,,
\eea
which holds identically for any $\tilde{\Xi}_0$. The coefficient for the $\mathcal{O}(\epsilon^{-2})$ term in $\mathcal{R}_2$ is therefore an exact constant of motion in the soft limit, the value of which is fixed entirely by the initial data; it cannot be dynamically sourced in the way argued in~\cite{Noisy}. 

Let us proceed to evaluate $\mathcal{F}$. Since it is conserved, we need only do so once. For the initial conditions, we require that there be no independent second-order mode present at early times $\eta \rightarrow 0$, but rather that any second-order fields be generated entirely by the convolution of first-order modes, consistent with the assumptions of~\cite{Noisy}. Mathematically this corresponds to demanding $G = D = 0$ in the homogeneous solution $P_{\rm hom} = G + D/\eta^3$. Noting that Eq.~\eqref{eq:P:eom} can equivalently be written
\bea
    \left( \eta^4 P' \right)' = \big( - \eta^3 \tilde{\Xi}_0 \big)' + 3 \eta^2 \tilde{\Xi}_0 \,,
\eea
the solution consistent with the initial conditions is
\bea
    \eta P' = - \tilde{\Xi}_0 + \frac{3}{\eta^3} \int_0^\eta d\bar{\eta} \, \bar{\eta}^2 \tilde{\Xi}_0(\bar{\eta}) \,.
\eea
Integrating once more, we are left with
\bea
    P = - \frac{1}{\eta^3} \int_0^\eta d\bar{\eta} \, \bar{\eta}^2 \tilde{\Xi}_0(\bar{\eta}) \,.
\eea
Substituting into Eq.~(\ref{eq:F:def}), one can check that the result vanishes identically, $\mathcal{F} = 0$. From Eq.~(\ref{eq:lim:is:F}) then
\bea\label{eq:lim:psi2:v2}
    \lim_{k \rightarrow 0} k^2 (\tilde{\psi}_2 - \mathcal{H} \tilde{v}_2 ) = 0 \,.
\eea

We emphasize that this result is not a property of the source $\tilde{\Xi}_0$, whose explicit form we have not used here. Rather it follows from the initial conditions (vanishing second-order homogeneous solution) together with the conservation law. Had we assumed a constant mode $G \neq 0$, $\mathcal{F}$ would have correspondingly been non-zero. However such a mode would have been a time-independent contribution $\tilde{\psi}_2 \supset G/k^2$ persisting on super-horizon scales, which would have had to be inserted by hand as an initial condition.

\subsection{Quadratic Completion}\label{sec:quad:completion}

All that remains to show is that $\mathcal{C}$ itself is free of poles. This quantity was derived in Appendix~\ref{sec:2nd:order:comoving:curvature}, and in radiation domination in Poisson gauge reads\footnote{This is the same object as in Eq.~(\ref{eq:C:main:text}) expressed in terms of $v_1$ for notational convenience.}
\bea\label{eq:Delta:quad:RD}
    \mathcal{C} = & - \mathcal{H}\nabla^{-2} \partial^i \bigg( - 6 \psi_1 \partial_i v_1 + v_1 \partial_i v_1' - v_1' \partial_i v_1 \bigg) \\
    & - \mathcal{H}^2 v_1^2 - \mathcal{H}v_1 v_1' + 2 v_1 \left( \psi_1' + 2 \mathcal{H} \psi_1 \right) + \frac{1}{2} (\partial_i v_1) (\partial^i v_1) - \frac{1}{2} \nabla^{-2} \partial_i \partial_j \left( \partial^i v_1 \partial^j v_1 \right) \,.
\eea
Straightforward power counting reveals that all the local terms in the second line are $\mathcal{O}(\epsilon^0)$ and so IR convergent. The last term in the second line involves an inverse Laplacian, but is also finite; in Fourier space, one can check that
\bea
    \nabla^{-2} \partial_i \partial_j \left( \partial^i v \partial^j v \right) \stackrel{\text{FT}}{\longrightarrow} - \frac{(\vec{k} \cdot \vec{q}\,) ( k^2 - \vec{k} \cdot \vec{q} \,)}{k^2} \tilde{v}_{\vec{q}} \, \tilde{v}_{\vec{k} - \vec{q}}
    \stackrel{k \rightarrow 0}{\longrightarrow} q^2 \mu^2 v_q v_q \,.
\eea
Meanwhile, the apparent pole in the first line is rendered finite by multiplying against the $\mathcal{O}(\epsilon)$ antisymmetric combination exactly as in Eq.~(\ref{eq:safe:pole}). The result is then that 
\bea\label{eq:C:limit}
    \mathcal{C} = \mathcal{O}(\epsilon^0) \,, \quad \text{such that} \quad \lim_{k \rightarrow 0} k^2 \mathcal{C} = 0 \,.
\eea
Comparing this behavior with the fact that $\lim_{k \rightarrow 0} \mathcal{Q} = \text{constant}$, as derived in Sec.~\ref{sec:RD:kurvature}, it is immediately clear that the two are manifestly unequal, $\nabla^2 \mathcal{C} \neq \mathcal{Q}$, and that their difference in the soft limit is precisely $\mathcal{Q}$ itself.

Finally, combining Eq.~(\ref{eq:C:limit}) with Eq.~(\ref{eq:lim:psi2:v2}), we see that
\bea\label{eq:lim:R2:0}
    \lim_{k \rightarrow 0} k^2 \mathcal{R}_2 = 0 \,.
\eea
Even though $\Delta \rho_2$ developed a white noise component, because at second order its relationship with $\mathcal{R}_2$ in Eq.~\eqref{eq:main:relation} is more complicated than the linear-level Poisson relationship~\eqref{eq:Noisy:relation} employed by~\cite{Noisy}, we see that $k^2 \mathcal{R}_2$ does not develop white noise. Instead, all the white noise in $\Delta \rho_2$ resides solely in $\mathcal{Q}$.

Before moving on, we comment that interestingly, the above calculation suggests that $\mathcal{R}_2$ itself should develop a white noise contribution at second order (though again, we emphasize that this is manifestly IR convergent; $\mathcal{R}_2$ does not develop an IR enhancement as suggested in~\cite{Noisy}). A quantitative characterization of this residual white noise in $\mathcal{R}_2$ is left to future work.

\section{Discussion \& Conclusions}\label{sec:discussion}

Let us review our main results. We have confirmed, by explicit calculation working to second order in cosmological perturbation theory, that the kurvature density $\Delta \rho$ develops a white noise spectrum on large scales: the quadratic source $\mathcal{Q}$ of Eq.~(\ref{eq:Q:def}) is finite and non-vanishing as $k \rightarrow 0$. We have computed this source explicitly for a perfect fluid during radiation domination, in doing so deriving the exact ``weight function'' introduced schematically in Ref.~\cite{Noisy}. In this sense, the LSWN effect argued for in Ref.~\cite{LSWN} is genuine for the kurvature density $\Delta \rho$.

What does \textit{not} follow from this, however, is the bound on the scale $k_{\rm BH}$, and by extension the small-scale power spectrum, argued for in Ref.~\cite{Noisy}. Here, the linear Poisson relation~\eqref{eq:Noisy:relation} was used to translate white noise in the second-order $\Delta \rho_2$ to white noise in $\nabla^2 \mathcal{R}_2$, resulting in a purported $1/k$ contribution to the full (dimensionless) power spectrum~\eqref{eq:bad:parameterization}. In the true second order relation~\eqref{eq:main:relation}, the white noise in $\Delta \rho_2$ resides solely in the quadratic term $\mathcal{Q}$. That is, the white noise in $\Delta \rho_2$ is \textit{not} inherited by $\nabla^2 \mathcal{R}_2$ in such a way as to give $\mathcal{P}_\mathcal{R}$ an IR enhancement. 

Moreover, by solving the second-order Einstein equations directly, we showed that $\lim_{k \rightarrow 0} \left( k^2 \mathcal{R}_2 \right) = 0$. This followed from a conservation law in the soft limit: the coefficient of the $\mathcal{O}(\epsilon^{-2})$ pole in $\mathcal{R}_2$---the quantity $\mathcal{F}$ defined in~\eqref{eq:F:def}---is an exact constant of motion during radiation domination. It cannot be dynamically sourced by quadratic combinations of first-order perturbations, contrary to the argument of~\cite{Noisy}; instead it retains its initial value, which is vanishing since a non-zero value would constitute a primordial second-order curvature mode on super-horizon scales. We remark that this result is robust; it follows directly from the structure of the evolution equation~\eqref{eq:P:eom} with appropriate boundary conditions, regardless of the explicit form of the quadratic sources $\Xi$.

Our proof limited itself to a perfect fluid in a radiation dominated era. While this was sufficient to identify the inconsistency in the bound of Ref.~\cite{Noisy}, which itself was derived under assumptions of a perfect fluid, it would be interesting to see what changes for an imperfect fluid like our own cosmological spacetime. For such a fluid, effects like viscosity, neutrino free streaming, and Silk damping can modify the shear $\sigma^2$ at small scales. This is important since $\sigma^2$ in a perfect radiation fluid is UV divergent for a scale-invariant primordial spectrum, as reflected in the growth of the weight function $\mathcal{W}$ in the right panel of Fig.~\ref{fig:W}. While the white noise in $\Delta \rho$ does not propagate to the CMB in the way that~\cite{Noisy} argued, this quantity is nevertheless observable in principle, being a local covariant scalar built from quantities measurable in the fluid rest frame, and so some resolution to this potential UV sensitivity would be satisfying. Finally, we have restricted to scalar perturbations, for which the vorticity contribution $\omega^2$ vanished in Eq.~(\ref{eq:kurvature:eom}). With the inclusion of vector and tensor modes, it will more generically be present. It would be interesting to re-examine the white noise in $\Delta \rho_2$ under these more realistic cosmological conditions. 

There are a number of other future directions that would be informative to pursue. As remarked at the end of Sec.~\ref{sec:quad:completion}, $\mathcal{R}_2$ itself develops a white noise contribution at second order coming from the quadratic completion term $\mathcal{C}$, which was $\mathcal{O}(\epsilon^0)$.\footnote{We emphasize again that this white noise is distinct from the IR enhancement claimed in~\cite{Noisy}, which would have corresponded to $\mathcal{O}(\epsilon^{-2})$ behavior in $\mathcal{C}$.} It would be interesting to characterize this numerically. Another future direction would be to establish the explicit mapping to CMB observables at second order. While at linear order the low multipole behavior is essentially just set by the Sachs-Wolfe effect, at higher order the relationship between $\mathcal{R}$ and CMB anisotropies is more complicated. In particular, mode coupling means that large angles are not purely soft $k$, and there exist many terms even at second order that could be important. While full second-order Boltzmann treatments exist~\cite{Bartolo:2006cu,Bartolo:2006fj}, the focus has typically been the bispectrum as the target observable, for which one can ignore the contribution from internal hard modes which is the origin of the white noise discussed here. Additionally, the full power spectrum at second order also receives a contribution coming from products of the form $\mathcal{R}_1 \mathcal{R}_3$, which to our knowledge has never been computed and is currently under investigation.

Finally, we found that the coefficient of the $\mathcal{O}(\epsilon^{-2})$ pole in $\mathcal{R}_2$---the quantity $\mathcal{F}$ defined in Eq.~(\ref{eq:F:def})---is a constant of motion in the soft limit. It would be good to understand better what this object is, and why it should be conserved. One possibility is it could be the soft limit of the Langlois-Vernizzi~\cite{Langlois:2005ii,Langlois:2005qp} charge, but this remains to be seen.\footnote{We thank Kylar Greene for making us aware of this conservation law.}

\bigskip
\bigskip
\noindent \textbf{Author's note:} While this work was being completed, we became aware that W. Hu had independently reached the same conclusion. We have coordinated posting of the two papers, but have refrained from reading one another's manuscripts in the interim so that the two analyses remain independent.

\begin{acknowledgments}
ANI is grateful to Gabriela Barenboim, Chris Byrnes, David Cyncynates, Yohei Ema, Josh Foster, Kylar Greene, Dan Hooper, Wayne Hu, Gordan Krnjaic, Albert Stebbins, and Vincent Vennin for valuable conversations and feedback. ANI is supported by NSF Grant PHY-2310429, Simons Investigator Award No.~824870, DOE HEP QuantISED award \#100495, the Gordon and Betty Moore Foundation Grant GBMF7946, and the U.S.~Department of Energy (DOE), Office of Science, National Quantum Information Science Research Centers, Superconducting Quantum Materials and Systems Center (SQMS) under contract No.~DEAC02-07CH11359. 
\end{acknowledgments}

\appendix
\section{Geometric Description of Spacelike Hypersurfaces}\label{sec:geometry}

Given a timelike unit 4-vector such as the fluid 4-velocity $u^\mu$, which satisfies $u^\mu u_\mu = -1$, one can describe the spatial hypersurfaces orthogonal to the flow in a fully covariant, geometric way. To do so, we first construct the spatial projection tensor $\mathsf{P}_{\mu \nu}$ as
\bea\label{eq:spatial:projector:def}
    \mathsf{P}_{\mu \nu} = g_{\mu \nu} + u_\mu u_\nu \,.
\eea
This object can be used to project quantities orthogonally to $u_\mu$, such that any tensor projected with $\mathsf{P}_{\mu \nu}$ is spatial. It satisfies 
\bea
    \mathsf{P}_{\mu \nu} u^\nu = 0 \,, \quad \mathsf{P}_{\mu}^\lambda \mathsf{P}_{\lambda \nu} = \mathsf{P}_{\mu \nu} \,.
\eea
The covariant derivative of $u_\mu$ describes how neighboring fluid worldlines move relative to one another and can be uniquely decomposed as~\cite{Wald:1984rg}
\bea\label{eq:fluid:decomposition}
    \nabla_\nu u_\mu = \frac{1}{3} \theta \mathsf{P}_{\mu \nu} + \sigma_{\mu \nu} + \omega_{\mu \nu} - a_\mu u_\nu \,.
\eea
Here, $\theta$ is the expansion scalar, defined by
\bea\label{eq:expansion:def}
    \theta = \nabla_\mu u^\mu \,,
\eea
which measures the isotropic expansion rate of neighboring fluid elements. Schematically for $\mathsf{V}$ a comoving volume element, $\theta = \dot{\mathsf{V}}/\mathsf{V}$. In an unperturbed FLRW background, then, $\theta = 3H$. The shear tensor $\sigma_{\mu \nu}$ is the trace-free, symmetric part of the decomposition~\eqref{eq:fluid:decomposition},
\bea\label{eq:shear:def}
    \sigma_{\mu \nu} = \frac{1}{2} \mathsf{P}_\mu^\alpha \mathsf{P}_\nu^\beta \big( \nabla_\beta u_\alpha + \nabla_\alpha u_\beta \big) - \frac{1}{3} \theta \mathsf{P}_{\mu \nu} \,,
\eea
which satisfies $\sigma_{\mu \nu} = \sigma_{\nu \mu}$, $\sigma_\mu^\mu = 0$, and $\sigma_{\mu \nu} u^\nu = 0$. Physically, it measures the distortion of fluid elements (without changing volume) due to anisotropic expansion. The vorticity tensor $\omega_{\mu \nu}$ is the anti-symmetric part of~\eqref{eq:fluid:decomposition},
\bea\label{eq:vorticity:def}
    \omega_{\mu \nu} = \frac{1}{2} \mathsf{P}_\mu^\alpha \mathsf{P}_\nu^\beta \big( \nabla_\beta u_\alpha - \nabla_\alpha u_\beta \big) \,,
\eea
which satisfies $\omega_{\mu \nu} = - \omega_{\nu \mu}$ and measures local rotation of the congruence. Finally, $a_\mu$ is the proper acceleration of fluid worldlines,
\bea\label{eq:acceleration:def}
    a_\mu = \dot{u}_\mu \,,
\eea
where $\dot{X} = u^\nu \nabla_\nu X$. Note that it is spatial, $a_\mu u^\mu = 0$. A vanishing acceleration $a_\mu = 0$ indicates geodesic flow while a non-zero value $a_\mu \neq 0$ reflects that gradients or forces are accelerating the fluid. 

The central equation governing the dynamical relationship among these quantities is the Raychaudhuri equation,
\bea
    \dot{\theta} = - \frac{1}{3} \theta^2 - 2\sigma^2 + 2\omega^2 + \nabla_\mu a^\mu - 4\pi G (\rho + 3 p) \,.
\eea
The minus sign accompanying the $4\pi G (\rho + 3p) \equiv R_{\mu \nu} u^\mu u^\nu$ term reflects the fact that ordinary matter focuses geodesics.\footnote{Here by ``ordinary'' matter we mean matter with $\rho + 3 p > 0$.} Similarly, the expansion has a self-focusing effect, and shear also enhances focusing. By contrast, rotation resists focusing. Finally, the acceleration term $\nabla_\mu a^\mu$ is not sign-definite; it can either be either focusing or defocusing since it is an externally supplied forcing term.\footnote{``External'' relative to the pure geometry of free fall.}

Now having defined these geometric quantities and established their kinematic and dynamical relationships, we are in a better position to understand the meaning of the kurvature density $\Delta \rho \equiv \rho - \theta^2/24\pi G$ of Eq.~(\ref{eq:kurvature:def}). Clearly this quantity measures the mismatch between the amount of matter present locally and how rapidly this local region is expanding. This mismatch is what produces intrinsic spatial curvature, and so $\Delta \rho$ is a fully covariant, geometric description of curvature inhomogeneities.

\section{Cosmological Perturbation Theory}\label{sec:2nd:order:quantities}

We follow the notation and conventions of Refs.~\cite{Malik:2008im,Christopherson:2011ra}. All tensorial quantities $T$ will be expanded to second order following the convention
\bea
    T(\eta,\bm{x}) = T_{(0)}(\eta) + \delta T_{(1)}(\eta,\bm{x}) + \frac{1}{2} \delta T_{(2)}(\eta,\bm{x}) \,,
\eea
where $(0)$ denotes a homogeneous background quantity and $(1)$, $(2)$ denotes the order of the inhomogeneous perturbation. The parenthesis around subscripts will be dropped when there is no danger of confusion with indices. The background will be taken to be the spatially flat Friedmann–Lema{\^i}tre–Robertson–Walker (FLRW) spacetime, with line element
\bea
    ds^2 = a^2 \big( - d\eta^2 + \delta_{ij} dx^i dx^j \big) \,,
\eea
with $\eta$ conformal time and $a(\eta)$ the scale factor.

\subsection{Metric Perturbations}\label{sec:metric}

We write the covariant metric as 
\bea
    g_{\mu \nu} = g_{\mu \nu}^{(0)} + \delta g_{\mu \nu}^{(1)} + \frac{1}{2} \delta g_{\mu \nu}^{(2)} \,,
\eea
with components
\begin{subequations}\label{eq:metric:covariant:components}
\begin{align}
    g_{00}^{(0)} & = - a^2 \,, 
    & \delta g_{00}^{(1)} & = - 2 a^2 \phi_1 \,, 
    & \delta g_{00}^{(2)} &= - 2 a^2 \phi_2 \,, \\
    g_{0i}^{(0)} & = 0 \,, 
    & \delta g_{0i}^{(1)} & = a^2 B_{1i} \,, 
    & \delta g_{0i}^{(2)} & = a^2 B_{2i} \,, \\
    g_{ij}^{(0)} & = a^2 \delta_{ij} \,, 
    & \delta g_{ij}^{(1)} & = 2 a^2 C_{1ij} \,, 
    & \delta g_{ij}^{(2)} & = 2 a^2 C_{2 ij} \,,
\end{align}
\end{subequations}
where at each order the $0i$ and $ij$ components can be decomposed into scalar, vector, and tensor parts
\bea
    B_i = \partial_i B - S_i \,,
\eea
\bea
    C_{ij} = - \psi \delta_{ij} + \partial_i \partial_j E + \frac{1}{2} \big( \partial_j F_i + \partial_i F_j \big) + \frac{1}{2} h_{ij} \,.
\eea
For future convenience, note that the trace of the latter reads
\bea
    C_i^i = - 3 \psi + \nabla^2 E \,.
\eea
The contravariant metric should similarly be expanded 
\bea
    g^{\mu \nu} = g^{\mu \nu}_{(0)} + \delta g^{\mu \nu}_{(1)} + \frac{1}{2} \delta g^{\mu \nu}_{(2)} \,.
\eea
The components, which can be found by demanding $g_{\mu \lambda} g^{\lambda \nu} = \delta_\mu^\nu$ at the desired order, read
\begin{subequations}
\begin{align}
    g^{00}_{(0)} & = - \frac{1}{a^2} \,,
    & \delta g^{00}_{(1)} & = \frac{2}{a^2} \phi_1 \,,
    & \delta g^{00}_{(2)} & = \frac{2}{a^2} \big( \phi_2 - 4 \phi_1^2 + B_{1i} B_1^i \big) \,,\\
    g^{0i}_{(0)} & = 0 \,,
    & \delta g^{0i}_{(1)} & = \frac{1}{a^2} B_1^i \,,
    & \delta g^{0i}_{(2)} & = \frac{1}{a^2} \big( B_2^i - 4 \phi_1 B_1^i - 4 B_{1j} C_1^{ij} \big) \,,\\
    g^{ij}_{(0)} & = \frac{1}{a^2} \delta^{ij} \,,
    & \delta g^{ij}_{(1)} & = - \frac{2}{a^2} C_1^{ij} \,,
    & \delta g^{ij}_{(2)} & = \frac{2}{a^2} \big( - C_2^{ij} + 4 C_1^{ik}C_{1k}^j - B_1^i B_1^j \big) \,.
\end{align}
\end{subequations}

\subsection{Connection Coefficients}\label{sec:connection}

The connection coefficients can be expanded as
\bea
    \Gamma^\alpha_{\beta \gamma} = {}_{(0)}{\Gamma}^\alpha_{\beta \gamma} + {}_{(1)}\delta \Gamma^\alpha_{\beta \gamma} + \frac{1}{2} {}_{(2)}\delta \Gamma^\alpha_{\beta \gamma} \,,
\eea
with the following entries:\\

\noindent$\mathbf{000}$\textbf{:}
\begin{subequations}
\bea
    {}_{(0)}\Gamma^0_{00} = \mathcal{H} \,,
\eea
\bea
    {}_{(1)}\delta\Gamma^0_{00} = \phi_1' \,,
\eea
\bea
    {}_{(2)}\delta\Gamma^0_{00} = \phi_2' - 4 \phi_1 \phi_1' + 2 B_1^i \big( B_{1i}' + \mathcal{H} B_{1i} + \partial_i \phi_{1} \big) \,,
\eea
\end{subequations}

\noindent$\mathbf{00i}$\textbf{:}
\begin{subequations}
\bea
    {}_{(0)}\Gamma^0_{0i} = 0 \,,
\eea
\bea
    {}_{(1)}\delta\Gamma^0_{0i} = \partial_i \phi_{1} + \mathcal{H} B_{1i} \,,
\eea
\bea
    {}_{(2)}\delta\Gamma^0_{0i} = \partial_i \phi_{2} + \mathcal{H} B_{2i} -4 \phi_1 \partial_i \phi_{1} - 4 \mathcal{H} \phi_1 B_{1i} + 2 B_1^j C_{1 ij}' + B_1^j \big(\partial_i B_{1j} - \partial_j B_{1i} \big) \,,
\eea
\end{subequations}

\noindent$\mathbf{0ij}$\textbf{:}
\begin{subequations}
\bea
    {}_{(0)}\Gamma^0_{ij} = \mathcal{H} \delta_{ij} \,,
\eea
\bea
    {}_{(1)}\delta\Gamma^0_{ij} = -2 \mathcal{H} \phi_1 \delta_{ij} - \frac{1}{2} \big( \partial_i B_{1j} + \partial_j B_{1i} \big) + C_{1 ij}' + 2 \mathcal{H} C_{1 ij} \,,
\eea
\bea
    {}_{(2)}\delta\Gamma^0_{ij} =& -2 \mathcal{H} \phi_2 \delta_{ij} - \frac{1}{2} \big( \partial_i B_{2j} + \partial_j B_{2i} \big) + C_{2 ij}' + 2 \mathcal{H} C_{2 ij} + 2 \mathcal{H} \big( 4 \phi_1^2 - B_{1k} B_1^k \big) \delta_{ij}\\
    & + 2 B_1^k \big( \partial_i C_{1 jk} + \partial_j C_{1 ik} - \partial_k C_{1 ij} \big) + 2 \phi_1 \big( \partial_j B_{1i} + \partial_i B_{1j} - 2 C_{1 ij}' - 4 \mathcal{H} C_{1 ij} \big) \,,
\eea
\end{subequations}

\noindent$\mathbf{i00}$\textbf{:}
\begin{subequations}
\bea
    {}_{(0)}\Gamma^i_{00} = 0 \,,
\eea
\bea
    {}_{(1)}\delta\Gamma^i_{00} = {B_1^i}' + \mathcal{H} B_1^i + \partial^i \phi_{1} \,,
\eea
\bea
    {}_{(2)}\delta\Gamma^i_{00} = {B_2^i}' + \mathcal{H} B_2^i + \partial^i \phi_{2} - 2 \phi_1' B_1^i - 4 C_1^{ij} \big( B_{1j}' + \mathcal{H} B_{1j} + \partial_j \phi_1 \big) \,,
\eea
\end{subequations}

\noindent$\mathbf{ij0}$\textbf{:}
\begin{subequations}
\bea
    {}_{(0)}\Gamma^i_{j0} = \mathcal{H} \delta^i_j \,,
\eea
\bea
    {}_{(1)}\delta\Gamma^i_{j0} = {C_{1j}^i}' + \frac{1}{2} \big( \partial_j B_1^i - \partial^i B_{1j} \big) \,,
\eea
\bea
    {}_{(2)}\delta\Gamma^i_{j0} = {C_{2j}^i}' + \frac{1}{2} \big( \partial_j B_2^i - \partial^i B_{2j} \big) - 4 C_1^{ik} C_{1 jk}' - 2 B_1^i \big( \mathcal{H} B_{1j} + \partial_j \phi_1 \big) + 2 C_1^{ik} \big( \partial_k B_{1j} - \partial_j B_{1k} \big) \,,
\eea
\end{subequations}

\noindent$\mathbf{ijk}$\textbf{:}
\begin{subequations}
\bea
    {}_{(0)}\Gamma^i_{jk} = 0 \,,
\eea
\bea
    {}_{(1)}\delta\Gamma^i_{jk} = \partial_j C_{1k}^i + \partial_k C_{1j}^i - \partial^i C_{1 jk} - \mathcal{H} B_1^i \delta_{jk} \,,
\eea
\bea
    {}_{(2)}\delta\Gamma^i_{jk} & = \partial_j C_{2k}^i + \partial_k C_{2j}^i - \partial^i C_{2 jk} - \mathcal{H} B_2^i \delta_{jk} + B_1^i \big( \partial_j B_{1k} + \partial_k B_{1j} \big) \\
    & + 2 B_1^i \big( 2 \mathcal{H} \phi_1 \delta_{jk} - C_{1 jk}' -2 \mathcal{H} C_{1 jk} \big) + 4 \mathcal{H} B_{1\ell} C_1^{i \ell} \delta_{jk} - 4 C_1^{i\ell} \big( \partial_j C_{1 k\ell} + \partial_k C_{1 j\ell} - \partial_\ell C_{1 jk} \big) \,.
\eea
\end{subequations}

\subsection{Matter Perturbations}\label{sec:matter}

We define the fluid 4-velocity $u^\mu$ to be the tangent vector field to the worldlines of fluid elements. Letting $x^\mu(\tau)$ be a fluid element's worldline, with $\tau$ the proper time comoving with the fluid, one has
\bea
    u^\mu = \frac{dx^\mu}{d\tau} \,.
\eea
From this definition along with the definition of proper time $d\tau^2 = - g_{\mu \nu} dx^\mu dx^\nu$, it follows that the fluid 4-velocity satisfies the constraint
\bea
    u^\mu u_\mu = - 1 \,.
\eea
Demanding this relation hold order by order, one can show that the components of $u^\mu$ are\footnote{Note that $u^\mu$ is fundamentally a quantity belonging to the matter sector. Because its normalization and coordinate components are tied to the metric, however, metric perturbations appear in its components.}
\begin{subequations}\label{eq:fluid:4velocity:components}
\bea
    u^0 = \frac{1}{a} \bigg( 1 - \phi_1 - \frac{1}{2} \phi_2 + \frac{3}{2} \phi_1^2 + \frac{1}{2} v_{1i} v_1^i + v_{1i} B_1^i \bigg) \,,
\eea
\bea
    u^i = \frac{1}{a} \bigg( v_1^i + \frac{1}{2} v_{2}^i \bigg) \,,
\eea
\end{subequations}
where we have introduced $v^i$ as the velocity perturbation of the fluid. Like the other vector quantities, it can be decomposed into a scalar and divergence-free vector part,
\bea
    v^i = \partial^i v + \mathfrak{v}^i \,,
\eea
where $\partial_i \mathfrak{v}^i = 0$. 

We expand the 4-velocity vector as
\bea
    u^\mu = u_{(0)}^\mu + \delta u_{(1)}^\mu + \frac{1}{2} \delta u_{(2)}^\mu \,.
\eea
Order-by-order, then, one has
\begin{subequations}\label{eq:u:contravariant:components}
\begin{align}
    u^0_{(0)} & = \frac{1}{a} \,,
    & \delta u^0_{(1)} & = - \frac{1}{a} \phi_1 \,,
    & \delta u^0_{(2)} & = - \frac{1}{a} \bigg( \phi_2 - 3 \phi_1^2 - v_{1i} v_1^i - 2 v_{1i} B_1^i \bigg) \,,\\
    u^i_{(0)} & = 0 \,,
    & \delta u^i_{(1)} & = \frac{1}{a} v_1^i \,,
    & \delta u^i_{(2)} & = \frac{1}{a} v_2^i \,.
\end{align}
\end{subequations}

One can obtain the covariant 4-velocity by using the metric to lower the index $u_\mu = g_{\mu \nu} u^\nu$. The components are
\begin{subequations}
\bea
    u_0 = - a \bigg( 1 + \phi_1 + \frac{1}{2} \phi_2 - \frac{1}{2} \phi_1^2 + \frac{1}{2} v_{1i} v_1^i \bigg) \,,
\eea
\bea
    u_i = a \bigg( v_{1 i} + B_{1i} + \frac{1}{2} \big( v_{2i} + B_{2i} \big) - \phi_1 B_{1i} + 2 C_{1 ij} v_1^j \bigg) \,.
\eea
\end{subequations}
Expanding
\bea
    u_\mu = u^{(0)}_\mu + \delta u^{(1)}_\mu + \frac{1}{2} \delta u^{(2)}_\mu \,,
\eea
the order-by-order components are
\begin{subequations}\label{eq:u:covariant:components}
\begin{align}
    u_0^{(0)} & = - a \,,
    & \delta u_0^{(1)} & = - a \phi_1 \,,
    & \delta u_0^{(2)} & = - a \big( \phi_2 - \phi_1^2 + v_{1i} v_1^i \big) \,,\\
    u_i^{(0)} & = 0 \,,
    & \delta u_i^{(1)} & = a \big( v_{1i} + B_{1i} \big) \,,
    & \delta u_i^{(2)} & = a \big( v_{2i} + B_{2i} - 2 \phi_1 B_{1i} + 4 C_{1 ij} v_1^j \big) \,.
\end{align}
\end{subequations}

The energy-momentum tensor of the fluid can be written in complete generality as
\bea
    T_{\mu \nu} = (\rho + p) u_\mu u_\nu + p g_{\mu \nu} + \pi_{\mu \nu} \,,
\eea
where, again, $\rho$ is the energy density and $p$ is the isotropic pressure. The anisotropic stress $\pi_{\mu \nu}$ is vanishing for a perfect fluid, but may in general be non-zero. Going forward, we will assume vanishing anisotropic stress, $\pi_{\mu \nu} = 0$, since the LSWN effect we are interested in is present even in its absence. 

\subsection{Geometric Quantities}\label{sec:geometric}

\noindent Here we derive and list the components of the spatial projection tensor, expansion, shear, vorticity, and acceleration corresponding to the matter 4-velocity $u^\mu$.\\

\noindent \textbf{Spatial Projection Tensor:}\\
\noindent Expanding Eq.~(\ref{eq:spatial:projector:def}) as 
\bea
    \mathsf{P}_{\mu \nu} = \mathsf{P}_{\mu \nu}^{(0)} + \delta \mathsf{P}_{\mu \nu}^{(1)} + \frac{1}{2} \delta \mathsf{P}_{\mu \nu}^{(2)} \,,
\eea
and using Eqs.~(\ref{eq:metric:covariant:components}) and (\ref{eq:u:covariant:components}), the components of the spatial projector at each order are
\begin{subequations}
\begin{align}
    \mathsf{P}_{00}^{(0)} & = 0 \,, 
    & \delta \mathsf{P}_{00}^{(1)} & = 0 \,, 
    & \delta \mathsf{P}_{00}^{(2)} &= 2 a^2 v_{1i} v_1^i \,, \\
    \mathsf{P}_{0i}^{(0)} & = 0 \,, 
    & \delta \mathsf{P}_{0i}^{(1)} & = - a^2 v_{1i} \,, 
    & \delta \mathsf{P}_{0i}^{(2)} & = - a^2 \big( v_{2i} + 2 \phi_1 v_{1i} + 4 C_{1 ij} v_1^j \big) \,, \\
    \mathsf{P}_{ij}^{(0)} & = a^2 \delta_{ij} \,, 
    & \delta \mathsf{P}_{ij}^{(1)} & =  2 a^2 C_{1 ij} \,, 
    & \delta \mathsf{P}_{ij}^{(2)} & = 2 a^2 C_{2 ij} + 2 a^2 \big( v_{1i} + B_{1i} \big) \big( v_{1j} + B_{1j} \big) \,.
\end{align}
\end{subequations}
For convenience, we also list the mixed index components at each order,
\begin{subequations}
\begin{align}
    {}_{(0)}\mathsf{P}_0^0 & = 0 \,, 
    & {}_{(1)}\delta \mathsf{P}_0^0 & = 0 \,, 
    & {}_{(2)}\delta \mathsf{P}_0^0 &= - 2 v_{1i} \big( v_1^i + B_1^i \big) \,, \\
    {}_{(0)}\mathsf{P}_i^0 & = 0 \,, 
    & {}_{(1)}\delta \mathsf{P}_i^0 & = v_{1i} + B_{1i} \,, 
    & {}_{(2)}\delta \mathsf{P}_i^0 &= v_{2i} + B_{2i} - 2 \phi_1 v_{1i} - 4 \phi_1 B_{1i} + 4 C_{1 ij} v_1^j \,, \\
    {}_{(0)}\mathsf{P}_0^i & = 0 \,, 
    & {}_{(1)}\delta \mathsf{P}_0^i & = - v_1^i \,, 
    & {}_{(2)}\delta \mathsf{P}_0^i &= - v_2^i - 2 \phi_1 v_1^i \,, \\
    {}_{(0)}\mathsf{P}_j^i & = \delta^i_j \,, 
    & {}_{(1)}\delta \mathsf{P}_j^i & = 0 \,, 
    & {}_{(2)}\delta \mathsf{P}_j^i &= 2 v_1^i \big( v_{1j} + B_{1j} \big) \,.
\end{align}
\end{subequations}\\

\noindent \textbf{Expansion:}\\
We expand the scalar expansion $\theta$ as
\bea
    \theta = \theta_0 + \delta \theta_1 + \frac{1}{2} \delta \theta_2 \,.
\eea
Using the definition~\eqref{eq:expansion:def} as well as the fluid velocity components in Eq.~(\ref{eq:u:contravariant:components}) and the connection components in Sec.~\ref{sec:connection}, we have
\begin{subequations}\label{eq:theta:components}
\bea
    \theta_0 = \frac{3 \mathcal{H}}{a} \,,
\eea
\bea
    \delta \theta_1 = \frac{1}{a} \big( - 3 \mathcal{H} \phi_1 + \nabla^2 v_1 + {C_{1i}^i}' \big) \,,
\eea
\bea
    \delta \theta_2 = \frac{1}{a} \bigg( & - 3 \mathcal{H} \phi_2 + \nabla^2 v_2 + {C_{2i}^i}' \\
    & + 9 \mathcal{H} \phi_1^2 + 2 \big( v_{1i}' + B_{1i}' \big) \big( v_1^i + B_1^i \big) + 3 \mathcal{H} v_{1i} \big( v_1^i + 2 B_1^i \big) \\
    & + 2 v_1^i \big( \partial_i \phi_1 + \partial_i C_{1 j}^j \big) + 2 C_1^{ij} \big( \partial_j B_{1i} - \partial_i B_{1j} \big) - 4 C_1^{ij} C_{1 ij}' - 2 \phi_1 {C_{1i}^i}' \bigg) \,,
\eea
\end{subequations}
where $\mathcal{H} = a H$.\\

\noindent \textbf{Shear:}\\
We expand the symmetric, traceless shear tensor $\sigma_{\mu \nu}$ as
\bea
    \sigma_{\mu \nu} = \sigma_{\mu \nu}^{(0)} + \delta \sigma_{\mu \nu}^{(1)} + \frac{1}{2} \delta \sigma_{\mu \nu}^{(2)} \,.
\eea
Note that in order to evaluate the kurvature density up to second order, one need only work out the shear up to first order, since what appears in this expression is $2 \delta \sigma_{1}^2 \equiv \delta \sigma_{(1)}^{\mu \nu} \delta \sigma_{\mu \nu}^{(1)}$. Also recall that the background value is vanishing, $\sigma_{\mu \nu}^{(0)} = 0$, and so one need only work out $\delta \sigma_{\mu \nu}^{(1)}$. Using the definition~\eqref{eq:shear:def} as well as the projection tensor and expansion components above, one can show
\begin{subequations}
\bea
    \delta \sigma_{00}^{(1)} = 0 \,,
\eea
\bea
    \delta \sigma_{0i}^{(1)} = 0 \,,
\eea
\bea
    \delta \sigma_{ij}^{(1)} = a \bigg( \frac{1}{2} \big( \partial_j v_{1i} + \partial_i v_{1j} \big) - \frac{1}{3} \nabla^2 v_1 \delta_{ij} + C_{1 ij}' - \frac{1}{3} {C_{1k}^k}' \delta_{ij} \bigg) \,.
\eea
\end{subequations}
The indices can be raised trivially with the background metric to obtain $\delta \sigma^{\mu \nu}_{(1)}$. The squared shear is then
\bea
    2 \delta \sigma_{1}^2 = \frac{1}{a^2} \bigg( {C_1^{ij}}' C_{1 ij}' - \frac{1}{3} & \big( {C_{1k}^k}' \big)^2 + \big( \partial_j v_{1i} + \partial_i v_{1j} \big) {C_1^{ij}}' - \frac{2}{3} \big( \nabla^2 v_1 \big) {C_{1k}^k}' \\
    & + \frac{1}{4} \big( \partial_j v_{1i} + \partial_i v_{1j} \big) \big( \partial^j v_1^i + \partial^i v_1^j \big) - \frac{1}{3} \big( \nabla^2 v_1 \big)^2 \bigg) \,.
\eea\\

\noindent \textbf{Vorticity:}\\
We expand the anti-symmetric vorticity $\omega_{\mu \nu}$ as
\bea
    \omega_{\mu \nu} = \omega_{\mu \nu}^{(0)} + \delta \omega_{\mu \nu}^{(1)} + \frac{1}{2} \delta \omega_{\mu \nu}^{(2)} \,.
\eea
As with the shear, we need only the first order perturbation, since the background quantity vanishes and what appears in Eq.~(\ref{eq:kurvature:eom}) is $2 \delta \omega_{1}^2 \equiv \delta \omega_{(1)}^{\mu \nu} \delta \omega_{\mu \nu}^{(1)}$. Evaluating the definition~\eqref{eq:vorticity:def} at this order we find
\begin{subequations}
\bea
    \delta \omega_{00}^{(1)} = 0 \,,
\eea
\bea
    \delta \omega_{0i}^{(1)} = 0 \,,
\eea
\bea
    \delta \omega_{ij}^{(1)} = \frac{a}{2} \bigg( \partial_j v_{1i} - \partial_i v_{1j} + \partial_j B_{1i} - \partial_i B_{1j} \bigg) \,.
\eea
\end{subequations}
One can trivially raise the indices with the background metric to obtain $\delta \omega^{ij}_{(1)}$, and so
\bea
    2 \delta \omega_{1}^2 = \frac{1}{4 a^2} \bigg( \partial_j v_{1i} - \partial_i v_{1j} + \partial_j B_{1i} - \partial_i B_{1j} \bigg) \bigg( \partial^j v_{1}^i - \partial^i v_{1}^j + \partial^j B_{1}^i - \partial^i B_{1}^j \bigg) \,.
\eea\\

\noindent \textbf{Acceleration:}\\
Finally, we expand the acceleration as
\bea
    a_\mu = a_\mu^{(0)} + \delta a_\mu^{(1)} + \frac{1}{2} \delta a_\mu^{(2)} \,,
\eea
where $a_\mu^{(0)} = 0$ in a flat FLRW background. Using the definition of Eq.~(\ref{eq:acceleration:def}) as well as the various quantities derived above, the components at first order are
\begin{subequations}\label{eq:a:covariant:components:1st}
\bea
    \delta a_0^{(1)} = 0 \,,
\eea
\bea
    \delta a_i^{(1)} = \partial_i \phi_1 + \big( v_{1i}' + B_{1i}' \big) + \mathcal{H} \big( v_{1i} + B_{1i} \big) \,,
\eea
\end{subequations}
and are second order
\begin{subequations}\label{eq:a:covariant:components:2nd}
\bea
    \delta a_0^{(2)} & = - 2 v_{1i} \bigg( \partial_i \phi_1 + \big( v_{1i}' + B_{1i}' \big) + \mathcal{H} \big( v_{1i} + B_{1i} \big) \bigg) \,,
\eea
\bea
    \delta a_i^{(2)} = & \partial_i \phi_2 + \big( v_{2i}' + B_{2i}' \big) + \mathcal{H} \big( v_{2i} + B_{2i} \big) \\
    & - 2 \phi_1 \big( v_{1i}' + \mathcal{H} v_{1i} \big) - 4 \phi_1 \big( B_{1i}' + \mathcal{H} B_{1i} \big) - 2 \phi_1' B_{1i} - 4 \phi_1 \partial_i \phi_1 \\
    & + 2 v_1^j \partial_j v_{1i} + 2 v_1^j \big( \partial_j B_{1i} - \partial_i B_{1j} \big) + 4 \big( C_{1 ij}' v_1^j + C_{1 ij} {v_1^j}' + \mathcal{H} C_{1 ij} v_1^j \big) \,.
\eea
\end{subequations}
The contravariant versions are
\begin{subequations}\label{eq:a:contravariant:components:1st}
\bea
    \delta a^0_{(1)} = 0 \,,
\eea
\bea
    \delta a^i_{(1)} = \frac{1}{a^2} \bigg( \partial^i \phi_1 + \big( {v_{1}^i}' + {B_{1}^i}' \big) + \mathcal{H} \big( v_{1}^i + B_{1}^i \big) \bigg) \,,
\eea
\end{subequations}
and
\begin{subequations}\label{eq:a:contravariant:components:2nd}
\bea
    \delta a^0_{(2)} & = \frac{2}{a^2} \big( v_1^i + B_1^i \big) \bigg( \partial_i \phi_1 + \big( v_{1i}' + B_{1i}' \big) + \mathcal{H} \big( v_{1i} + B_{1i} \big) \bigg) \,,
\eea
\bea
    \delta a_i^{(2)} = \frac{1}{a^2} \bigg( & \partial^i \phi_2 + \big( {v_{2}^i}' + {B_{2}^i}' \big) + \mathcal{H} \big( v_{2}^i + B_{2}^i \big) \\
    & - 2 \phi_1 \big( {v_{1}^i}' + \mathcal{H} v_{1}^i \big) - 4 \phi_1 \big( {B_{1}^i}' + \mathcal{H} B_{1}^i \big) - 2 \phi_1' B_{1}^i - 4 \phi_1 \partial^i \phi_1 \\
    & + 2 v_1^j \partial_j v_{1}^i + 2 v_1^j \big( \partial_j B_{1}^i - \partial^i B_{1j} \big) + 4 \big( {C_{1}^{ij}}' v_{1j} - C_{1}^{ij} B_{1j}' - \mathcal{H} C_{1}^{ij} B_{1j} - C_1^{ij} \partial_j \phi_1 \big) \bigg) \,.
\eea
\end{subequations}
The divergence of the acceleration is then
\bea\label{eq:1st:order:div:acceleration}
    \big(\nabla_\mu a^\mu \big)_{(1)} = \frac{1}{a^2} \bigg( \nabla^2 \phi_1 + \nabla^2 \big( v_1' + B_1' \big) + \mathcal{H} \nabla^2 \big( v_1 + B_1 \big) \bigg) \,.
\eea
We omit the expression for $\big(\nabla_\mu a^\mu \big)_{(2)}$ since it is sufficiently long so as to be impractical. It can be straightforwardly assembled from the pieces above as $\big( \nabla_\mu a^\mu \big)_{(2)} = \partial_\mu \delta a_{(2)}^\mu + {}_{(0)}\Gamma^\mu_{\mu \nu} \delta a_{(2)}^\nu + 2 {}_{(1)}\Gamma^\mu_{\mu \nu} \delta a_{(1)}^\nu$.

\section{Dynamics and Constraints}\label{sec:dynamics:constraints}

The metric and matter sectors described above are related by the Einstein field equations
\bea\label{eq:Einstein}
    G_{\mu \nu} = 8 \pi G T_{\mu \nu} \,,
\eea
where $G_{\mu \nu} = R_{\mu \nu} - \frac{1}{2} R g_{\mu \nu}$, with $R_{\mu \nu}$ the Ricci tensor and $R = g^{\mu \nu} R_{\mu \nu}$ the Ricci scalar. In general, one can project~\eqref{eq:Einstein} onto components tangent and orthogonal to the fluid 4-velocity $u^\mu$ in order to obtain two constraint equations --- the ``energy'' and ``momentum'' constraints. These must be satisfied by the initial data at each instant. The remaining components of the Einstein equation contain two dynamical equations. Finally, the Bianchi identity $\nabla_\mu G^\mu_\nu = 0$ enforces energy-momentum conservation
\bea\label{eq:energy:momentum:conservation}
    \nabla_\mu T^{\mu \nu} = 0 \,,
\eea
which can also be split into two conservation equations for energy and momentum.

\subsection{Background}

At the background level, the momentum constraint is trivially satisfied and the energy constraint is simply the first Friedmann equation,
\bea\label{eq:Friedmann:1:app}
    \mathcal{H}^2 = \frac{8 \pi G}{3} a^2 \rho_0 \,,
\eea
where $\mathcal{H} = a H$ is the conformal Hubble rate. Similarly, there is only one independent evolution equation --- the second Friedmann equation, 
\bea\label{eq:Friedmann:2:app}
    \mathcal{H}' = - \frac{4 \pi G}{3} a^2 \big(\rho_0 + 3 p_0 \big) \,,
\eea
where $'$ denotes a derivative with respect to conformal time. Finally, from Eq.~(\ref{eq:energy:momentum:conservation}) one obtains the continuity equation, 
\bea\label{eq:continuity:app}
    \rho'_{0} + 3 \mathcal{H} \big(\rho_0 + p_0 \big) = 0 \,,
\eea
with momentum conservation equation being trivially satisfied. Note that of Eqs.~(\ref{eq:Friedmann:1:app}), (\ref{eq:Friedmann:2:app}), and (\ref{eq:continuity:app}), only two are independent.

\subsection{First Order}

At first order, scalar, vector, and tensor perturbations decouple, and so one can derive the constraint and dynamical equations for the various sectors independently. For the scalar perturbations, the energy and momentum constraints are
\bea
    3 \mathcal{H} \big( \psi_1' + \mathcal{H} \phi_1 \big) - \nabla^2 \psi_1 - \mathcal{H} \nabla^2 \big( E_1' - B_1 \big) = - 4 \pi G a^2 \delta \rho_1 \,,
\eea
\bea
    \psi_1' + \mathcal{H} \phi_1 = - 4 \pi G a^2 \big( \rho_0 + p_0 \big) \big(v_{1} + B_{1} \big) \,.
\eea
The two dynamical equations at this order are\footnote{Keep in mind that had we not set the anisotropic stress to zero, there would be additional source terms on the right-hand side.}
\bea\label{eq:1st:order:psi:eom}
    \psi_1'' + 2 \mathcal{H} \psi_1' + \mathcal{H} \phi_1' + \big( 2 \mathcal{H}' + \mathcal{H}^2 \big) \phi_1 = 4 \pi G a^2 \delta p_1 \,,
\eea
\bea\label{eq:1st:order:scalar:shear:eom}
    E_1'' - B_1' + 2 \mathcal{H} \big( E_1' - B_1 \big) + \psi_1 - \phi_1 = 0 \,.
\eea
Finally, energy and momentum conservation give
\bea\label{eq:1st:order:energy:conservation}
    \delta \rho_1' + 3 \mathcal{H} \big(\delta \rho_1 + \delta p_1 \big) = \big( \rho_0 + p_0 \big) \big( 3 \psi_1' - \nabla^2 E_1' - \nabla^2 v_1 \big) \,,
\eea
\bea\label{eq:1st:order:momentum:conservation}
    v_1' + B_1' + \big(1 - 3 c_s^2 \big) \mathcal{H} \big( v_1 + B_1 \big) + \phi_1 + \frac{\delta p_1}{\rho_0 + p_0} = 0 \,,
\eea
where $c_s^2 \equiv p_0'/\rho_0'$ is the adiabatic speed of sound. We do not reproduce the vector and tensor sector equations here, as our eventual focus will be scalar perturbations.

\subsection{Second Order}

Here we collect energy-momentum conservation and Einstein equations at second order. From the former, we have the energy conservation equation
\bea\label{eq:2nd:order:energy:conservation}
    \delta \rho_2' & + 3 \mathcal{H} \left(\delta \rho_2 + \delta P_2 \right) + (1+w) \rho_0 \left( {C_{2i}^i}' + \partial^i v_{2i} \right) + 2 v_1^i \partial_i \left( \delta \rho_1 + \delta P_1 \right) + 2 \left( \delta \rho_1 + \delta P_1 \right) \left( {C_{1i}^i}' + \partial^i v_{1i} \right) \\ 
    & + 2 (1+w) \rho_0 \bigg( \left( 2 v_1^i + B_1^i \right) \left( v_{1i}' + B_{1i}' \right) + \phi_1 \partial_i v_1^i - 2 C_{1 ij}' C_1^{ij} + v_1^i\partial_i  \left( C_{1j}^j + 2 \phi_1 \right) + 4 \mathcal{H} v_1^i \left(2 v_{1i} + B_{1i} \right) \bigg) = 0 \,,
\eea
and the momentum conservation equation
\bea\label{eq:2nd:order:momentum:conservation}
    (1+w) & \bigg( \rho_0' \left( v_{2i} + B_{2i} \right) + \rho_0 \left( v_{2i}' + B_{2i}' \right) + \rho_0 \left( \partial_i \phi_2 + 4 \mathcal{H} \left( v_{2i} + B_{2i} \right) \right) \bigg) + \partial_i \delta P_2 \\
    & + 2 \left( \delta \rho_1' + \delta P_1' \right) \left( v_{1i} + B_{1i} \right) + 2 \left( \delta \rho_1 + \delta P_1 \right) \left( v_{1i}' + B_{1i}' \right) + 2 \left( \delta \rho_1 + \delta P_1 \right) \left( \partial_i \phi_1 + 4 \mathcal{H} (v_{1i} + B_{1i}) \right) \\
    & - 2 (1+w) \rho_0' \left( \left( v_{1i} + 2 B_{1i} \right) \phi_1 - 2 C_{1 ij} v_1^j \right) \\
    & + 2 (1+w) \rho_0 \bigg( (v_{1i} + B_{1i}) \left( {C_{1j}^j}' + \partial_j v_1^j \right) - B_{1i} \left( \phi_1' + 8 \mathcal{H} \phi_1 \right) + 2 C_{1ij}' v_1^j + 2 C_{1i}^j v_{1j}' \bigg) \\
    & + 2 (1+w) \rho_0 \bigg( v_1^j \left( \partial_j (v_{1i} + B_{1i}) - \partial_i B_{1j} + 8 \mathcal{H} C_{1 ij} \right) - \phi_1 \left( v_{1i}' + 2 B_{1i}' + 2 \partial_i \phi_1 + 4 \mathcal{H} v_{1i} \right) \bigg) = 0 \,.
\eea
Meanwhile, the $00$ component of the Einstein equation gives the Hamiltonian constraint 
\bea\label{eq:2nd:order:energy:constraint}
    \nabla^2 & C_{2i}^i - \partial^i \partial^j C_{2 ij} + 2 \mathcal{H} \big(3 \mathcal{H} \phi_2 + \partial_i B_2^i - {C_{2i}^i}' \big) + \partial_i C_{1j}^j \big( \partial^i C_{1j}^j - 4 \partial_j C_1^{ij} \big) \\
    & + 2 B_1^i \bigg( \partial_i {C_{1j}^j}' - \partial^j C_{1 ij}' + \frac{1}{2} \big( \nabla^2 B_{1i} - \partial_i \partial_j B_1^j \big) - 2 \mathcal{H} \big( \partial_i \phi_1 + 2 \partial^j C_{1 ij} - \partial_i C_{1j}^j \big) \bigg) \\
    & + 4 C_1^{ij} \bigg( 2 \partial_i \partial_k C_{1 j}^k - \partial_i \partial_j C_{1k}^k - \nabla^2 C_{1 ij} + 2\mathcal{H} \big( C_{1 ij}' - \partial_j B_{1i} \big) \bigg) \\
    & + \partial_i C_{1 jk} \big( 2 \partial^j C_1^{ik} - 3 \partial^i C_1^{jk} \big) + 4 \partial_i C_1^{ij} \partial^k C_{1 jk} + C_{1 ij}' \big( {C_1^{ij}}' - 2 \partial^i B_1^j \big) \\
    & + {C_{1i}^i}' \big( 2 \nabla^2 B_1 - {C_{1j}^j}' + 8 \mathcal{H} \phi_1 \big) + \frac{1}{2} \partial_i B_{1j} \big( \partial^j B_1^i + \partial^i B_1^j \big) - 6 \mathcal{H}^2 \big(4 \phi_1^2 - B_{1i} B_1^i \big) - (\nabla^2 B_1)^2 - 8 \mathcal{H} \phi_1 \nabla^2 B_1 \\
    & = - 8 \pi G a^2 \bigg( \delta \rho_2 + 2 \rho_0 (1+w) v_{1i} \big( v_1^i + B_1^i \big) \bigg) \,,
\eea
while the $0i$ component gives the momentum constraint
\bea\label{eq:2nd:order:momentum:constraint}
    \partial_i {C^{k}_{2k}}' & -\partial^k {C_{2ik}'} -\frac{1}{2}\left(\partial_i\partial^k B_{2k}-\nabla^2 B_{2i}\right) -2\mathcal{H}\partial_i \phi_{2} +16\mathcal{H} \partial_i \phi_{1}\phi_1 - 2{C^{j}_{1j}}' \partial_i \phi_{1}\\
    & +2 C'_{1ij}\left(2 \partial_k C_{1}^{kj} - \partial^j C_{1k}^k +\partial^j \phi_{1} \right) +4C_{1}^{kj}\left( \partial_j C'_{1ik} - \partial_i C'_{1jk} + \frac{1}{2}\partial_j \left(\partial_i B_{1k} -\partial_k B_{1i}\right) \right)\\
    & +2B_{1}^{j} \bigg(\partial_i \partial^k C_{1kj} -\partial_i \partial_j C_{1 k}^{k} + \partial^k \partial^j C_{1ik} -\nabla^2 C_{1ij} -2\mathcal{H} \partial_i B_{1j} \bigg) -\left(\partial_j B_{1i}+ \partial_i B_{1j}\right) \partial^j \phi_{1} \\
    & +2\left(\partial_j B_{1i} -\partial_i B_{1j}\right) \left( \frac{1}{2} \partial^j C_{1k}^k - \partial_k C_{1}^{jk} \right) - 2 \partial_j C_{1ik} \left(\partial^j B_{1}^{k} -\partial^k B_{1}^{j} \right) + 2 \left(\partial_j B_{1}^{j} \right) \partial_i \phi_{1}\\
    & + 2\phi_1 \bigg( \partial_i \partial^j B_{1j} - \nabla^2 B_{1i} + 2 \left( \partial^j C'_{1ij} - \partial_i {C_{1j}^j}' \right) \bigg) -2C_{1jk}' \partial_i C^{jk}\\ & = 16\pi G \left( \frac{1}{2} \left(v_{2i} + B_{2i} \right) - \phi_1 \left(v_{1i} + 2 B_{1i} \right) +2C_{1ik} v_{1}^{k} +\left(\delta\rho_1+\delta P_1\right)\left(v_{1i} + B_{1i} \right) \right) \,.
\eea
The $ij$ component is sufficiently complicated that we omit it here, opting to just write the more specialized form in Eq.~(\ref{eq:2nd:order:Einstein:ij:longitudinal}) below.\\

\noindent \textbf{Scalar sector in Poisson gauge:}\\
These expressions simplify considerably upon restricting to scalar perturbations and fixing the gauge. As in the main text, we work in Poisson gauge. Setting $E = B = 0$ and also using that $\psi_1 = \phi_1$ at first order, Eq.~(\ref{eq:2nd:order:energy:conservation}) simplifies to
\bea
    \delta \rho_2' & + 3 \mathcal{H} \left(\delta \rho_2 + \delta P_2 \right) + (1+w) \rho_0 \left( \nabla^2 v_2 - 3 \psi_2' \right) + 2 (\partial^i v_1) \partial_i \left( \delta \rho_1 + \delta P_1 \right) + 2 \left( \delta \rho_1 + \delta P_1 \right) \left( \nabla^2 v_1 - 3 \psi_1' \right) \\ 
    & + 2 (1+w) \rho_0 \bigg( 2 (\partial^i v_1) (\partial_i v_{1}') + 8 \mathcal{H} (\partial^i v_1) (\partial_i v_1) + 3 \psi_1' \psi_1 + \psi_1 \nabla^2 v_1 - (\partial^i v_1)(\partial_i \psi_1) \bigg) = 0 \,,
\eea
and (\ref{eq:2nd:order:momentum:conservation}) simplifies to
\bea
    (1+w) & \rho_0 \left( \frac{\rho_0'}{\rho_0} \partial_i v_2 + \partial_i v_2' + \partial_i \phi_2 + 4 \mathcal{H} \partial_i v_2 \right) + \partial_i \delta P_2 + 2 \left( \delta \rho_1' + \delta P_1' \right) \partial_i v_1 \\
    & + 2 \left( \delta \rho_1 + \delta P_1 \right) \left( \partial_i v_1' + \partial_i \phi_1 + 4 \mathcal{H} \partial_i v_1 \right) - 6 (1+w) \rho_0' \psi_1 \partial_i v_1 \\
    & + 2 (1+w) \rho_0 \bigg( (\partial_i v_1) \nabla^2 v_1 + (\partial^j v_1) \partial_i \partial_j v_1 - 5 \psi_1' \partial_i v_1 - 3 \psi_1 \partial_i v_1' - 2 \psi_1 \partial_i \psi_1 - 12 \mathcal{H} \psi_1 \partial_i v_1 \bigg) = 0 \,.
\eea
For the Einstein equations, Eq.~(\ref{eq:2nd:order:energy:constraint}) simplifies to
\bea\label{eq:2nd:order:energy:constraint:longitudinal}
    \nabla^2 \psi_2 - 3 \mathcal{H} \big( \psi_2' + \mathcal{H} \phi_2 \big) + 12 \mathcal{H}^2 \psi_1^2 + 3 {\psi_1'}^2 + 8 \psi_1 \nabla^2 \psi_1 + 3 \big( \partial_i \psi_1 \big) \big( \partial^i \psi_1 \big) = 4 \pi G a^2 \bigg(\! \delta \rho_2 + 2 \rho_0 (1+w) \big( \partial_i v_1 \big) \big( \partial^i v_1 \big) \!\bigg) \,,
\eea
while Eq.~(\ref{eq:2nd:order:momentum:constraint}) simplifies to
\bea\label{eq:2nd:order:momentum:constraint:longitudinal}
    \partial_i \left( \psi_2' + \mathcal{H} \phi_2 \right) + 2 \psi_1' \partial_i \psi_1 - 8 \mathcal{H} \psi_1 \partial_i \psi_1 = - 4 \pi G a^2 \bigg( \rho_0 (1+w) \left( \partial_i v_2 - 6 \psi_1 \partial_i v_1 \right) + 2 \left( \delta \rho_1 + \delta P_1 \right) \partial_i v_1 \bigg) \,.
\eea
Finally, the trace of the $ij$ Einstein equation becomes
\bea\label{eq:2nd:order:Einstein:ij:longitudinal}
    3 & \psi_2'' + 3 \mathcal{H} \left(2 \psi_2' + \phi_2' \right) + \nabla^2 (\phi_2 - \psi_2) + 3 \left( 2 \frac{a''}{a} - \mathcal{H}^2 \right) \phi_2
    \\
    & - 6 (\partial^i \psi_1) (\partial_i \psi_1) - 24 \mathcal{H} \psi_1 \psi_1' - 8 \psi_1 \nabla^2 \psi_1 - 3 \psi_1' \psi_1' + 12 \psi_1^2 \left( \mathcal{H}^2 - 2 \frac{a''}{a} \right) \\
    & \quad\quad\quad\quad\quad\quad\quad\quad\quad\quad\quad\quad = 4\pi G a^2 \bigg( 3 \delta P_2 + 2 \rho_0 (1+w) (\partial_i v_1) (\partial^i v_1) \bigg) \,.
\eea

\section{Second Order Comoving Curvature Perturbation}\label{sec:2nd:order:comoving:curvature}

Here we review the construction of the gauge invariant comoving curvature perturbation at second order, $\mathcal{R}_2 \equiv \bar{\psi}_2^{\rm co}$, where we use a bar to denote the gauge transformation induced in a tensorial quantity
\bea
    \bar{T} = e^{\mathcal{L}_\xi} T \,,
\eea
with $\mathcal{L}_\xi$ a Lie derivative with respect to the gauge generator vector $\xi^\mu$. Up to second order, 
\bea
    \bar{\delta T}_2 = \delta T_2 + \mathcal{L}_{\xi_2} T_0 + \mathcal{L}_{\xi_1}^2 T_0 + 2 \mathcal{L}_{\xi_1} \delta T_1 \,,
\eea
where we have similarly expanded $\xi^\mu = \xi_1^\mu + \frac{1}{2} \xi_2^\mu$. In general, we write
\bea
    \xi^\mu = (\alpha, \partial^i \beta + \gamma^i) \,,
\eea
with $\alpha = \alpha_1 + \frac{1}{2} \alpha_2$ and similarly for $\beta$, $\gamma^i$. One can check that at second order, the metric quantity $\psi_2$ transforms as
\bea\label{eq:psi2:bar:general}
    \bar{\psi}_2 = \psi_2 - \mathcal{H} \alpha_2 - \frac{1}{4} \chi_i^i + \frac{1}{4} \nabla^{-2} \partial_i \partial_j \chi^{ij} \,,
\eea
with $\chi_{ij}$ the combination defined in~\cite{Malik:2008im}
\bea
    \chi_{ij} = & 2 \bigg[ \left(\mathcal{H}^2+\frac{a''}{a}\right) \alpha_1^2 + \mathcal{H} \left( \alpha_1 \alpha_1' + \partial_k \alpha_1 \xi_1^k \right) \bigg] \delta_{ij} \\
    & + 4 \left( \alpha_1 \left( C_{1 ij}' + 2 \mathcal{H} C_{1 ij} \right) + \partial_k C_{1 ij} \xi_1^k + C_{1 ik} \partial_j \xi_1^k + C_{1 jk} \partial_i \xi_1^k \right) + 2 \left( B_{1 i} \partial_j \alpha_1 + B_{1j} \partial_i \alpha_1 \right) \\
    & + 4 \mathcal{H} \alpha_1 \left( \partial_j \xi_{1 i} + \partial_i \xi_{1 j} \right) - 2 \partial_i \alpha_1 \partial_j \alpha_1 + 2 (\partial_i \xi_{1 k}) (\partial_j \xi_1^k) + \alpha_1 \left( \partial_j \xi_{1i}' + \partial_i \xi_{1j}' \right) + \xi_1^k \partial_k \left( \partial_j \xi_{1i} + \partial_i \xi_{1j} \right) \\
    & + (\partial_k \xi_{1 i}) (\partial_j \xi_1^k) + (\partial_k \xi_{1j})(\partial_i \xi_1^k) + \xi_{1i}' \partial_j \alpha_1 + \xi_{1j}' \partial_i \alpha_1 \,.
\eea
In comoving gauge with slicing $\bar{v}_1 + \bar{B}_1 = 0$ and isotropic threading $\bar{E}_1 = 0$, the first order gauge generator is particularly simple: 
\bea
    \alpha_1^{\rm co} = v_1 + B_1 \,, \quad \beta_1^{\rm co} = \gamma_{1i}^{\rm co} = 0 \,.
\eea
Expressing the result in the Poisson gauge employed in the main text, we have
\bea
    \chi_{ij}^{\rm co}\big|_{\rm P} = 2 \bigg( \left( \mathcal{H}^2 + \frac{a''}{a} \right) v_1^2 + \mathcal{H}v_1 v_1' - 2 v_1 \left( \psi_1' + 2 \mathcal{H} \psi_1 \right) \bigg) \delta_{ij} - 2 \partial_i v_1 \partial_j v_1 \,.
\eea
The second order scalar part of the gauge generator $\alpha_2$ in comoving gauge reads
\bea
    \alpha_2 = v_2 + \nabla^{-2} \partial^i \bigg( - 6 \psi_1 \partial_i v_1 + v_1 \partial_i v_1' - v_1' \partial_i v_1 \bigg) \,.
\eea
Combining the above in Eq.~(\ref{eq:psi2:bar:general}), we obtain for the gauge invariant comoving curvature perturbation at second order (evaluated in Poisson gauge),
\bea\label{eq:R2:appendix}
    \mathcal{R}_2 = \psi_2 - \mathcal{H} v_2 + \mathcal{C} \,,
\eea
where we have defined the quadratic completion
\bea\label{eq:Delta:quad}
    \mathcal{C} = & - \mathcal{H}\nabla^{-2} \partial^i \bigg( - 6 \psi_1 \partial_i v_1 + v_1 \partial_i v_1' - v_1' \partial_i v_1 \bigg) \\
    & - \left( \mathcal{H}^2 + \frac{a''}{a} \right) v_1^2 - \mathcal{H}v_1 v_1' + 2 v_1 \left( \psi_1' + 2 \mathcal{H} \psi_1 \right) + \frac{1}{2} (\partial_i v_1) (\partial^i v_1) - \frac{1}{2} \nabla^{-2} \partial_i \partial_j \left( \partial^i v_1 \partial^j v_1 \right) \,.
\eea

\section{Quadratic Sources}\label{sec:quad:sources}

The quadratic sources appearing in the second-order Einstein equations~\eqref{eq:EE1}, \eqref{eq:EE2}, \eqref{eq:EE3} and energy-momentum conservation equations~\eqref{eq:EM1}, \eqref{eq:EM2} are
\begin{subequations}\label{eq:quad:sources}
\bea
    Q^{(1)} = 12 \mathcal{H}^2 \psi_1^2 + 3 (\psi_1')^2 + 8 \psi_1 \nabla^2 \psi_1 + 3 (\partial_i \psi_1)(\partial^i \psi_1) - 3 \mathcal{H}^2 (1+w) (\partial_i v_1)(\partial^i v_1) \,,
\eea
\bea
    Q^{(2)}_i = 2 \psi_1' \partial_i \psi_1 - 8 \mathcal{H} \psi_1 \partial_i \psi_1 - 24 \pi G a^2 \rho_0 (1+w) \psi_1 \partial_i v_1 + 8 \pi G a^2 (1+c_s^2) \delta \rho_1 \partial_i v_1 \,,
\eea
\bea
    Q^{(3)} = - 6 (\partial_i \psi_1)(\partial^i \psi_1) - 24 \mathcal{H} \psi_1 \psi_1' - 8 \psi_1 \nabla^2 \psi_1 - 3 (\psi_1')^2 - 12 \left( 2 \frac{a''}{a} - \mathcal{H}^2 \right) \psi_1^2 - 8 \pi G a^2 \rho_0 (1+w) (\partial_i v_1)(\partial^i v_1) \,,
\eea
\bea
    Q^{(4)} = &2 (1+c_s^2) \bigg( (\partial^i v_1) \partial_i \delta \rho_1 + \delta\rho_1 (\nabla^2 v_1 - 3 \psi_1') \bigg) \\
    & + 2 (1+w) \rho_0 \bigg( 2 (\partial_i v_1) (\partial^i v_1') + 8 \mathcal{H} (\partial_i v_1)(\partial^i v_1) + 3 \psi_1' \psi_1 + \psi_1 \nabla^2 v_1 - (\partial^i v_1) (\partial_i \psi_1) \bigg) \,,
\eea
\bea
    Q^{(5)}_i = & 2 (1+c_s^2) \delta \rho_1 \bigg( \frac{\delta \rho_1'}{\delta \rho_1} \partial_i v_1 + \partial_i v_1' + \partial_i \phi_1 + 4 \mathcal{H} \partial_i v_1 \bigg) - 6 (1+w) \rho_0' \psi_1 \partial_i v_1 \\
    & + 2 (1+w) \rho_0 \bigg( (\partial_i v_1) \nabla^2 v_1 + (\partial^j v_1) \partial_i \partial_j v_1 - 5 \psi_1' \partial_i v_1 - 3 \psi_1 \partial_i v_1' - 2 \psi_1 \partial_i \psi_1 - 12 \mathcal{H} \psi_1 \partial_i v_1 \bigg) \,.
\eea
\end{subequations}

\bibliography{biblio}

\end{document}